\documentclass[twocolumn]{aastex631}
\usepackage{amsthm,amsmath,amsfonts,amssymb}
\usepackage[normalem]{ulem}

\usepackage{mathtools,breqn,comment,graphicx}
\usepackage{xcolor,epstopdf}% To incorporate .eps illustrations using PDFLaTeX, etc.
\usepackage[caption=false]{subfig}% Support for small, `sub' figures and tables
\theoremstyle{plain}% Theorem-like structures provided by amsthm.sty
\theoremstyle{definition}

\theoremstyle{remark}

\usepackage{amsmath,amsthm,mathrsfs}
\usepackage{graphicx,psfrag,epsf}
\usepackage{enumerate}
\usepackage{natbib}
\usepackage{url} % not crucial - just used below for the URL 
\usepackage{amssymb,amsmath}
\usepackage{color,xcolor,nicefrac}

\def\xmm{\textit{XMM--Newton}}
\def\1es{1ES~1553+113}

\def\apj{Astrophys. J.}
\def\jcap{JCAP}
\def\cmin{$C_{\mathrm{min}}$}

\def\cminValEq{c_{\mathrm{min}}}
\def\cminVal{$\cminValEq$}

\def\cmineq{C_{\mathrm{min}}}

\def\cminsys{$C_{\mathrm{min,sys}}$}
\def\cminsyseq{C_{\mathrm{min,sys}}}
\newcommand\Var{\text{Var}}
\renewcommand\skew{\text{skew}}
\newcommand\kurt{\text{kurt}}

\newcommand\E{\text{E}}

\def\chimin{$\chi^2_{\mathrm{min}}$}
\def\Poiss{\mathrm{Poisson}}
\def\NB{NB}
\def\PIG{P-IG}

\def\ml{maximum--likelihood}

\def\LHat{$\mathcal{L}(\hat{\theta})$}

\def\DP{$D_P$}
\def\Int{\mathrm{int}}

\def\sigmaInt{$\sigma_{\mathrm{int}}$}
\def\sigmaIntiEq{\sigma_{\mathrm{int},i}}
\def\sigmaIntjEq{\sigma_{\mathrm{int},j}}
\def\sigmaIntRelEq{\sigma_{\mathrm{int},i}/\hat{\mu}_i}
\def\sigmaIntRel{$\sigma_{\mathrm{int},i}/\hat{\mu}_i$}

\def\gof{goodness--of--fit}
\shorttitle{Poisson regression with systematic errors}
\shortauthors{Bonamente et al.}

\begin{document}
%\articletype{original research paper}
%\begin{frontmatter}
%%%%%%%%%%%%%%%%%%%%%%%%%%%%%%%%%%%%%%%%%%%%%%
%%                                          %%
%% Enter the title of your article here     %%
%%                                          %%
%%%%%%%%%%%%%%%%%%%%%%%%%%%%%%%%%%%%%%%%%%%%%%
  \title{Maximum--likelihood regression with systematic errors for astronomy and the physical sciences: \\I. Methodology and goodness--of--fit statistic of Poisson data}
%\title{A sample article title with some additional note\thanksref{T1}}
%\runtitle{Poisson regression with systematic errors}
%\thankstext{T1}{A sample of additional note to the title.}

%\begin{aug}
\author[0000-0002-8597-9742]{Massimiliano Bonamente}\affiliation{Department of Physics and Astronomy, University of Alabama in Huntsville, Huntsville, AL 35899}
\author[0000-0002-9516-8134]{Yang Chen}\affiliation{Department of Statistics, University of Michigan, Ann Arbor, MI 48109}
\author[0000-0003-1212-4089]{Dale  Zimmerman}\affiliation{Department of Statistics and Actuarial Science,
University of Iowa, Iowa City, IA 52242}
%\textsuperscript{a}\thanks{CONTACT M. Bonamente. Email: bonamem@uah.edu},
%Yang Chen\textsuperscript{b} and Dale  Zimmerman\textsuperscript{c}}
%\affil{\textsuperscript{a} Department of Physics and Astronomy, University of Alabama in %Huntsville, Huntsville, AL 35899;\\ 
%\textsuperscript{b} Department of Statistics, University of Michigan, Ann Arbor, MI 48109; \\
%\textsuperscript{c} Department of Statistics and Actuarial Science,
%University of Iowa, Iowa City, IA 52242}
%}
%%%%%%%%%%%%%%%%%%%%%%%%%%%%%%%%%%%%%%%%%%%%%%%
%% Only one address is permitted per author. %%
%% Only division, organization and e-mail is %%
%% included in the address.                  %%
%% Additional information can be included in %%
%% the Acknowledgments section if necessary. %%
%%%%%%%%%%%%%%%%%%%%%%%%%%%%%%%%%%%%%%%%%%%%%%%
%\author[A]{\fnms{Massimiliano} \snm{Bonamente}\ead[label=e1]{bonamem@uah.edu}}
%\and
%\author[B]{\fnms{Dale} \snm{Zimmerman} \ead[label=e2]{dale-zimmerman@uiowa.edu}}
%\author[B]{\fnms{???} \snm{???}\ead[label=e2,mark]{???@???}}
%\and
%\author[B]{\fnms{???} \snm{???}\ead[label=e3,mark]{???@???}}
%%%%%%%%%%%%%%%%%%%%%%%%%%%%%%%%%%%%%%%%%%%%%%
%% Addresses                                %%
%%%%%%%%%%%%%%%%%%%%%%%%%%%%%%%%%%%%%%%%%%%%%%
%\address[A]{Department of Physics and Astronomy,
%University of Alabama in Huntsville, Huntsville AL 35899 (U.S.A), \printead{e1}}
%\address[B]{}
%\address[B]{???, \printead{e2,e3}}
%\end{aug}
%\maketitle
\begin{abstract}
The paper presents a new statistical method that enables the use
 of systematic errors in the \ml\ regression of
integer--count Poisson data to a parametric model. The method is primarily aimed at the characterization
of the \gof\ statistic in the presence of the over--dispersion that is induced by sources of
systematic error, and is based on a {quasi-\ml} method that
retains the Poisson distribution of the data.
We show that the Poisson deviance, which is the usual \gof\ statistic and that is
commonly referred to in astronomy as the \emph{Cash} statistics, can
be easily generalized in the presence of systematic errors, under rather general conditions.
The method and the associated statistics are first developed theoretically, and
then they are tested with the aid of numerical simulations
and further illustrated with real--life data from astronomical observations. The statistical methods 
presented in this paper are intended as a simple
general--purpose framework to include additional sources of uncertainty for the analysis of
integer--count data in a variety of practical data analysis situations.
\end{abstract}

%\begin{keywords}
%\kwd{Probability distributions}
%\kwd{Maximum--likelihood methods}
%\kwd{Poisson distribution}
%\kwd{Poisson log--likelihood}
%\kwd{Parameter estimation}
% \kwd{systematic errors}
\keywords{Astrostatistics(1882); Regression(1914); Maximum likelihood estimation(1901); Poisson distribution(1898); Parametric hypothesis tests(1904); Measurement error model(1946)}
%\end{keywords}

%\end{frontmatter}

\section{Introduction: Systematic Errors and Poisson Regression}
\label{sec:intro}

It is common practice, in the physical sciences and related disciplines, to 
classify uncertainties or `errors' in quantities of interest into two categories:
`random' or `statistical' errors on one hand, and `systematic' errors on the other. The two types of error
are generally attributed to the following assumptions: (1) the method of measurement of the random variable yields an inherent
sampling distribution of measurements, when the experiment is repeated under the same experimental
conditions (e.g., a normal or Poisson distribution); and (2) that there may be unknown sources of uncertainty, either in the measurement 
process itself or in the underlying theory, which may cause additional errors that do not average down with
repeated measurements. 

This subject is of practical importance for the measurement
of physical quantities {\color{blue} \citep[e.g.,][]{ku1969, ISO2008,NIST1994}.}
While `random' errors have been the purview of probability and statistics since its 
beginnings, i.e. via the characterization of a random variable by its sampling distribution
and moments, there seems to be no general consensus as to what is meant by 'systematic' errors, and how to provide a probabilistic model for them. 
 For example, comments on this subject by \citealt{glosup1996}, \citealt{eisenhart1962}, \citealt{dorsey1953} or in Sec.~5.3 of \citealt{jeffreys1966},
suggest the presence of one or more additional random variables
besides those of interest to the experimenter. These variables can be included in the statistical
modelling but considered uninteresting (i.e., a \emph{nuisance} parameter such as a background level,
e.g.,  
\citealt{lyons2008, lyons2020, cowan2019}), or 
not be accounted in the experimental design altogether (i.e, a \emph{hidden} or \emph{latent} variables,
e.g., \citealt{skrondal2004}). In both cases, these variables
have the potential to affect the measurement
of the variable(s) of interest by biasing the the estimation
of the mean of target variables of interest, and also contributing additional variance (and possibly
higher--order moments). This interpretation guides the model for 
systematic errors to be presented in this paper.

A review by \cite{vandyk2023} describes several methods that are commonly used in the physical sciences
to address sources of systematic error. Some of the simplest and more common methods
use a one--parameter--at--a--time error propagation, or
several parameters simultaneously \citep[see, e.g.][]{heinrich2007}. These methods are especially common,
and have been the \emph{de facto} standard for astrophysics \citep[see, e.g., notable applications by][]{allen2004, freedman2001}.
More complex frequentist and Bayesian methods are also available \citep[e.g.][]{acero2022, lee2011, cousins1992}, depending on  the data and type of application (i.e., point estimation, interval estimation, upper limits, goodness of fit, etc.). Certain such methods require subsidiary measurements to constrain 
the likelihood associated with the nuisance parameters, and to marginalize the posterior for the parameters of interest with respect to the nuisance parameters, e.g., see application to
cosmological parameters by \cite{planck2016-cosmology,planck2014-likelihood}, or for the 
calibration of X--ray instruments by \cite{lee2011,drake2006}.

In regression applications  with normal data,
it is customary
to treat such systematic errors with the addition of a fiducial systematic variance 
to that of the data, 
assuming  independence
between measurement and systematic errors.
This is the usual method
to address sources of unknown systematic error in many physics and astronomy regression
applications \citep[e.g.][]{bevington2003,nordin2008}, especially when there is no additional
information on the physical origin of the systematic error. A popular
method for astronomers to include systematic errors in the linear regression is the
BCES (Bivariate Correlated Errors and intrinsic Scatter) method
 of \cite{akritas1996}. That method, however, only applies to the linear regression, and does not
 explicitly provide a \ml\ \gof\ statistic.

For Poisson--distributed count data, on the other hand, the fixed relationship between the mean and variance of the distribution complicates the statistical treatment of systematic errors.
Overdispersed count--data distributions, such as the negative binomial 
or the Poisson inverse Gaussian (\PIG), are a possible solution
to the problem \citep[see, e.g.,][]{hilbe2011,hilbe2014, cameron2013}. However, 
it is often desirable to maintain the Poisson distribution of the data, 
both because of its simpler log--likelihood properties 
and the belief that the data  do
result from a fixed--rate Poisson process.
Motivated by this challenge, a new method to deal with  the regression of Poisson
 data with systematic errors was proposed by \cite{bonamente2023}, which is based on the modelling of systematic errors via
an intrinsic model variance, instead of using an overdispersed model
for the data themselves. 
Based on that idea, this paper presents in more detail the statistical framework 
for the regression of Poisson data in the presence of systematic errors, and certain
asymptotic results on the Poisson deviance as its goodness--of--fit distribution.

This paper is structured as follows: Sec.~\ref{sec:MLPoisson} provides a review of
\ml\ regression with Poisson data and what is known
concerning its \gof\ statistic. Sec.~\ref{sec:model} presents the details of the model for
systematic errors for Poisson data, followed by the new proposed method of regression and
its \gof\ statistic in Sec.~\ref{sec:cmin}. The model is then tested in Sec.~\ref{sec:cminSym} with the aid of numerical
simulations,  Sec.~\ref{sec:applications} describes 
methods of use of this model, and
Sec.~\ref{sec:conclusions} presents our 
conclusions.

\section{The maximum--likelihood Poisson regression} 
\label{sec:MLPoisson}
This section describes the data model and the goodness--of--fit statistic
that results from the maximum--likelihood method of regression with Poisson data.
Although other methods of regression are  available, such as the least--squares method, 
the (generalized) method of moments 
\citep[for a review, see e.g.][]{cameron2013}
or the method of maximum entropy \citep[e.g.][]{jaynes1957}, the \ml\ criterion 
\citep[e.g.][]{fisher1925} is commonly used  in the physical sciences, and it is the method considered in this study.

\subsection{The data model}
\label{sec:PoissonDataModel}
Regression of cross--sectional Poisson data via the
\ml\ method uses 
a data model of the type $(x_i, y_i)$, for $ i=1,\dots,N$ independent
 measurements $y_i\sim \Poiss(\mu_i)$ at different values of independent variables $x_i$.
The \ml\ method returns the Poisson log--likelihood \LHat\ as the fit statistic, 
\begin{equation}
\begin{aligned}
    \mathcal{L}(\hat{\theta}) & \coloneqq \ln L(\hat{\theta}) = \ln \left( \prod_{i=1}^N \dfrac{e^{\displaystyle- \hat{\mu}_i} \hat{\mu}_i^{\displaystyle y_i}}{y_i!} \right) = \\
     & \sum_{i=1}^N \left( - \hat{\mu}_i + y_i \ln \hat{\mu}_i - \ln y_i! \right)
    \end{aligned}
    \label{eq:PoissonL}
\end{equation}
where $\theta=(\theta_1,\dots,\theta_m)$ are $m$ adjustable parameters of a model $y=f(x;\theta)$ 
that can be non--linear,
and $\hat{\mu}=(\hat{\mu}_1,\dots,\hat{\mu}_N)$ are the means of the Poisson distributions evaluated at the best--fit parameter values $\hat{\theta}$, i.e., $\hat{\mu}_i= f(x_i;\hat{\theta})$. 
For a textbook review of Poisson \ml\ regression, see e.g. \cite{cameron2013, james2006,bonamente2022book}.

The  $x$ variable, with its $x_i$ fixed positions,
represents the predictor or independent variable,
and the response or dependent variable $y_i$ represent an integer number of occurrences
detected as a function of the predictor.
This data model applies to a variety of data in the physical sciences and related disciplines,
and it is also commonly used in other fields such as econometrics \citep{cameron2013}, where
the most common model used is the simple linear model with $m=2$ parameters.
In this paper, the term "model" is used to indicate interchangeably what astronomers might refer to as either a physical model (e.g., the type of radiation emitted by an astronomical source), or  a detector model that characterizes the method of detection and how
counts are collected by the instrument. The term is therefore used in this paper in a  
general sense to indicate a functional form for the parent mean $\mu_i$ for the 
Poisson distribution of detected counts, according to \eqref{eq:PoissonL}.

In certain disciplines, it is common to use several predictors
for the response, e.g., $x \in \mathbb{R}^k$ with $k \geq2$,  typically using a linear model 
with $m=2$ \citep[i.e., the multiple linear regression, e.g.][]{rao1973, cameron2013} or generalized linear models \citep[GLM, e.g.][]{mccullagh1989}. 
In applications for astronomy and the physical sciences, however, emphasis is placed on a physically--motivated and often non--linear model $y(x)$ that is a function of just one predictor, $k=1$.
In astronomy,
a scalar $x$ often represents either time or the energy of a particle or a photon, and 
the associated data are referred to respectively as time series (or light curves) and spectra \citep[see, e.g.][]{feigelson2012}.
The class of generalized linear models with intercept, $\mu_i=\eta(X_i\, \theta^T)$ with
$X_i \in \mathbb{R}^{m+1}$ and
$\eta(\cdot)$ the link function \citep[e.g.][]{mccullagh1989}, is included
in the general class of functions $f(x, \theta)$ considered in this paper. 
 The results to be presented in this paper apply to predictors with any dimension, $k \geq 1$,
although examples and simulations will only be reported for scalar predictors with $x \in \mathbb{R}$.

\subsection{The Poisson deviance}
The Poisson {deviance} is defined as twice the difference between the maximum achievable log--likelihood $\mathcal{L}(y)$ and that of the fitted model $\mathcal{L}(\hat{\theta})$,
\begin{equation}
  D_P \coloneqq 2 (\mathcal{L}(y)-\mathcal{L}(\hat{\theta})) = 2 \sum_{i=1}^N \left( y_i \ln \left(\dfrac{y_i}{\hat{\mu}_i}\right) - (y_i-\hat{\mu}_i) \right)
  \label{eq:DP}
\end{equation}
and it is the goodness--of--fit statistic of choice for the Poisson regression, with $\hat{\mu}_i=\mu_i(\hat{\theta})$ the best--fit Poisson means obtained for the best--fit parameters $\hat{\theta}$ \citep[e.g.][]{cameron2013}. The Poisson deviance differs from
\eqref{eq:PoissonL} via the subtraction of $\mathcal{L}(y)$, which is the 
likelihood for the \emph{saturated} model (i.e., the model in which the $N$ means are equal to the
measured data) and it is not a function of the model parameters.
The Poisson deviance statistic is also referred to as the \emph{G--statistic} \citep{bishop1975}.~\footnote{
See also Sec.~5.3.2 of \cite{cameron2013}, where the factor of 2 was omitted.
}

This statistic has received much attention for the \gof\ assessment 
of high--energy astronomy count data.
In this field, the \DP\ statistic is known as the \emph{Cash} or \emph{C statistic} and it is 
commonly referred to as \cmin, after the pioneering work of X--ray astronomer W.~Cash and others
\citep{cash1976, cash1979,baker1984}. The \emph{Cash} statistic
is also the fit statistic of choice for
the major high--energy astronomy spectral packages \citep{kaastra1996,arnaud1996, doe2007}.~\footnote{In \texttt{XSPEC},
the \emph{Cash} statistic in \eqref{eq:DP} is reported also as \texttt{cstat} and attributed to J. Castor in \cite{arnaud2011}.}
According to \eqref{eq:DP}, we define the function
\begin{equation}
    C(\mu) \coloneqq 2 \sum\limits_{i=1}^N \left( y_i \ln \left(\dfrac{y_i}{\mu_i}\right) - (y_i-\mu_i) \right) = \sum\limits_{i=1}^N C^{(i)}(\mu)
  \label{eq:cmin}
\end{equation}
with $\cmineq=C(\hat{\mu})$ the usual Poisson deviance evaluated at the ML estimates of the Poisson means. 
The definition \eqref{eq:cmin} implies 
that $y_i \sim \Poiss(\mu_i)$ be independent
measurements, as per the data model of Sec.~\ref{sec:PoissonDataModel}.

\subsection{Distribution of the Poisson deviance} 
\label{sec:DP}

According to \eqref{eq:DP}, the Poisson deviance $D_P$ is a \emph{likelihood ratio} statistic,
as already pointed out by \cite{cash1979}. 
A likelihood ratio statistic, as originally proposed by \cite{neyman1928a}, is generally defined
by the ratio of two likelihoods,
        \begin{equation}
        \Lambda=\frac{\sup\limits_{\boldsymbol{\theta}_0} L(\theta|y)}
            {\sup\limits_{\boldsymbol{\theta}} L(\theta|y)},
            \label{eq:Lambda}
    \end{equation}
with $\boldsymbol{\theta} \subset \mathbb{R}^m$ the full
parameter space and $\boldsymbol{\theta}_0 \subset \boldsymbol{\theta}$ a subset of parameter space representing the null hypothesis.
A general result for the asymptotic distribution of certain likelihood--ratio statistics is 
provided by the \emph{Wilks theorem} \citep{wilks1938,wilks1962}~\footnote{See also Sec.~30.3 of \cite{cramer1946} or Ch.~6 of \cite{rao1973}}. For a simple null hypothesis in which all of the $m$ parameters are fixed, the Wilks
theorem may be reported as
\begin{equation}
  -2 \ln \Lambda^{(m)} = -2 (\mathcal{L}(\theta_0^{(m)}) - \mathcal{L}(\hat{\theta})) \overset{a}{\sim}  \chi^2(m)
  \label{eq:wilks1}
\end{equation}
 in the asymptotic limit of a large sample. In Eq.~\ref{eq:wilks1}, $\mathcal{L}(\theta_0^{(m)})$ is the log--likelihood for a simple null hypothesis $H_0$
represented by
a set of fixed (or true) parameters $\theta=\theta_0$, with $\theta_0 \in \boldsymbol{\theta}_0$;
and $\mathcal{L}(\hat{\theta})$ is the usual log--likelihood obtained by using the $m$ best--fit maximum--likelihood parameters that were left free in the fit. 

The Wilks theorem also applies to a composite hypothesis with $s < m$ of the parameters fixed
at a reference (or true) value, and the remaining $m-s$ parameters 
being free to adjust to their best--fit value.
In that case, the Wilks theorem results in
\begin{equation}
  -2 \ln \Lambda^{(m-s)} = -2 (\mathcal{L}(\theta_0^{(m-s)}) - \mathcal{L}(\hat{\theta})) \overset{a}{\sim}  \chi^2(m-s)
  \label{eq:wilks2}
\end{equation}
One of the key hypotheses for the applicability of the Wilks theorem is that partial derivatives of the Poisson log--likelihood with respect 
to  $\theta_i$ are continuous \citep[see e.g.,][Sec. 5f]{rao1973}.

Within the context of likelihood ratios for $D_P$, the term
$\mathcal{L}(y)$ used in \eqref{eq:DP} 
represents the log--likelihood for the \emph{saturated} or \emph{perfect} model, as described earlier.
Unfortunately, the hypotheses of Wilks' theorem are \emph{not} satisfied when the 
logarithm of the denominator of \eqref{eq:Lambda} is  $\mathcal{L}(y)$, as is the case for  
the Poisson deviance defined by \eqref{eq:DP}. This is because the perfect 
or saturated model in which $\hat{\mu}_i=y_i$ is not described by the same continuous parameterization $\theta$ as the null model (term $\mathcal{L}(\hat{\theta})$).
Therefore the Wilks theorem does not apply in general to the Poisson deviance $D_P$, and thus
it is {not} possible to assume that $D_P \sim \chi^2(N-m)$ in general, for all values of the parent mean and for any form of the null--hypothesis parametric model.

Asymptotic and approximate distributions for $D_P=\cmineq$ are available in the literature.
\cite{mccullagh1986} found the conditional distribution of $D_P$ for 
log--linear models in the extensive and sparse regime (i.e., large $N$ and small $\mu$) via the method
of cumulants \citep{mccullagh1984}, including first-- and second--order corrections
to the asymptotic $\chi^2$ distribution that applies to the large--mean regime. Similar results were 
also obtained by \cite{kaastra2017} and \cite{bonamente2020}, however limited to 
the case of a simple hypothesis. 
An in--depth analysis of the Poisson deviance, including more accurate approximations that generalize the
\cite{mccullagh1986} results to a broader class of models, is presented in \cite{li2024}.
A useful approximation to the distribution of the Poisson deviance 
is available in the large--mean regime, where $D_P$ is approximately distributed as 
$\chi^2(N-m)$, due to the approximation of the Poisson distribution to a normal
distribution in that regime \citep[e.g., in the \emph{Gaussian limit} of][]{cash1979}.
In the astrophysics literature, the large--mean regime is usually interpreted
as approximately $\geq 10$ or 20 counts per bin \citep[e.g., see discussion in Ch. 15 of][]{bonamente2022book}.

These complications are in contrast with the case of normal data, where
the maximum--likelihood \gof\ statistic
\begin{equation}
    S = \sum\limits_{i=1}^N \dfrac{(y_i - \hat{\mu}_i)^2}{\sigma^2_i}
\end{equation}
is distributed as $S \sim \chi^2(N-m)$  for a variety of 
models and situations, under rather general conditions \citep[e.g.,\emph{Cram\'{e}r's theorem,}][]{cramer1946,greenwood1996}; for this reason, this statistic is usually referred to
as the minimum chi--squared statistics (\chimin) in astronomical applications. 
This simple result for normal data is one of the main reasons why some practitioners 
use the minimum chi--squared statistic for parameter estimation
and hypothesis testing of Poisson data, 
even though such method leads to biased results
\citep[e.g. ][]{humphrey2009}.

\section{A model for the regression of Poisson data with systematic errors}
\label{sec:model}

This section provides the details of the proposed method of analysis, starting with a definition
of the type of systematic errors that are being considered.

\subsection{Systematic errors as source of overdispersion}
\label{sec:sysErrDefinition}

In the development of a statistical model of systematic errors, we take the narrow
view that systematics only provide an additional source of \emph{variance} beyond that implied by the
distribution of the data, without affecting the mean. While this assumption may seem restrictive, 
consider that any
systematic shift or bias in the measurement of $y_i$ can in principle be accounted by means of
suitable model components that contribute to the model function $f(x_i;\theta)$, provided they
are a reasonably smooth function of the regressor variable $x$. 

Practically, this view of systematic errors means that if the
response variable $y_i$ is affected by 
systematic errors, the
data analyst has corrected for the \emph{mean} effect resulting from those errors. This is 
in fact what is routinely
done in  certain
physics and astronomy applications, where such sources of systematics
as instrumental gain or calibration are corrected with fixed factors that aim to 
provide a mean adjustment to the model \citep[see, e.g., the calibration efforts for X--ray astronomy by][]{tsujimoto2011, plucinsky2017}.
From an experimental
viewpoint, this process results in a best--fit model that generally follows the data without
systematic departures (in the mean), yet with a larger--than--expected variance.

In this work, we consider strictly independent measurements $y_i$, as already emphasized in Sec.~\ref{sec:PoissonDataModel}; this is a key assumption of the method. In certain applications, data may be binned with a finer resolution than the effective detector resolution, thereby introducing a certain degree of correlation among the detected counts in adjacent bins, and therefore also among the systematics. The choice of bin size is primarily a data
reduction problem that is not directly addressed by this work \citep[see, e.g.][]{kaastra2016}, and therefore it is assumed that bin--by--bin independence is satisfied. Nonetheless, it is important to point out that calibration, and therefore
correction--in--the--mean, are often carried out over a range of the independent
variable that covers several bins simultaneously \citep[e.g.][for X--ray astronomy examples]{drake2006, lee2011}. This is likely to introduce a degree of correlation in the systematics between adjacent bins that is not accounted in this model of systematic errors, which would require additional considerations.

A way of specifying the effect of systematic errors on the response $y_i$ is illustrated via the compounding 
%overdispersion 
representation of random variables,~\footnote{Compounding is also referred to as \emph{mixing} in the statistical literature. Compounding formulas for mean and variance are reported, for example, 
in Sec.~4.3 of \cite{mood1974}, or Prob. 23, Chapter~V of \cite{feller-vol-2}. See App.~\ref{app:distributions}
for a brief review.} with a non--negative distribution 
of choice (such as the gamma) for the Poisson mean  $\mu_i$, leading to overdispersion. 
No restriction is placed on the origin of such overdispersion, i.e., 
on the shape of the
compounding distribution for $\mu_i$.
In fact, we intend to
develop a  general--purpose 
model that enables (a) an estimate of the systematic
variance, when the use of an \emph{a priori} value is impractical; (b) a method of regression for Poisson count data 
in the presence of these errors; and, critically,  (c) 
the evaluation of the distribution for the revised fit statistic (i.e., a deviance $D_P$ in the presence
of systematics)  for the purpose of hypothesis testing. These are the main goals of the method 
to be presented in this paper.

\subsection{Review of models for overdispersion in count data}
\label{sec:overdispersionReview}

Some of the more common distributions to model overdispersed integer data are the
negative binomial (\NB) distribution, which results from the compounding of a Poisson distribution
with a gamma--distributed rate parameter \citep[e.g.][]{greenwood1920}, or the
Poisson--inverse--Gaussian (\PIG) distribution \citep[e.g.][]{tweedie1957, sichel1971, atkinson1982}.
Such compounding (or mixing) models
have been used in a variety of applications across
the sciences, see e.g. \cite{karlis2005} or \cite{willmot1986} for a review of 
Poisson mixing distributions and applications. In 
addition to the gamma and inverse Gaussian distributions, some of the other notable
mixing distributions include the lognormal distribution \citep{bulmer1974} and the truncated normal
distribution \citep{patil1964}. The \NB\ distribution is commonly used in GLM models
\citep[e.g., the NB1 and NB2 parameterizations][]{cameron2013,mccullagh1989}.

A method to account for the additional variance induced by systematic errors is therefore
the \emph{direct} use of one such compounded Poisson distribution for the data, in place of the default Poisson model of Sec.~\ref{sec:PoissonDataModel}. There are two major reasons, however, to seek an alternate route for certain applications.
First, there is substantial added computational
cost in the use of alternative distributions such as the \NB\ or the \PIG. For example,
the \PIG\ distribution makes use of Bessel functions of the second kind \citep[e.g.][]{holla1967,atkinson1982}; likewise, the likelihood associated with the
\NB\ is more complex than for the Poisson distribution \citep[see, e.g.][]{hilbe2011}.
More critical, however, is the fact that the \ml\ goodness--of--fit statistic becomes practically intractable,
even for the relatively less complex \NB\ model \citep[see, e.g.][Ch.~4]{cameron2013}.
The situation would therefore become even more dire
than for the {\color{blue} relatively} simple Poisson regression statistic \DP,
which itself lacks the
simple distributional properties of the chi--squared statistic for normal data 
{\color{blue} (see  Sec.~\ref{sec:DP})}.

On the other hand, and moving away for a moment from `classical' compounded distributions, one could follow purely Bayesian methods for the inclusion of systematic errors
in the analysis. Such methods would typically start with the choice of a prior distribution $p(\mu_i$) 
to evaluate the
posterior distributions for $\mu_i$.
In this case, 
 assessment of the 
goodness--of--fit would be addressed via Bayes factors
or by means of information criteria such as AIC \citep[Akaike Information Crierion,][]{akaike1974},
BIC \citep[Bayesian Information Criterion][]{schwarz1978} or  similar \citep[for a review, see e.g.][]{stoica2004, gelman2013}.
Numerical evaluations are typically required for the posterior distribution, of a complexity that is
probably of similar order to that of the numerical minimization of the \DP\ statistic for the classical approach, depending on a number of factors such as the complexity of the model and that of the prior
distribution of choice.
The main drawback, however, is that the the Bayesian approach does not have a simple goodness--of--fit
statistic to directly test the null hypothesis $H_0$ of a true model. 

Certain applications may not require
an accurate determination of the distribution of the 
goodness of fit statistic. In those cases, the \NB\ or \PIG\ models or other compounded--variable variations,
or alternatively Bayesian methods, may be the 
most appropriate route to model overdispersion. However, most physics and astronomy applications do require it,  in that it is usually critical to perform an accurate test of a complex hypothesis as a means of
validation or rejection of an underlying multi--parameter theory.
Examples in X--ray astronomy include  most cosmological studies such as the measurement of
the Hubble constant \citep[e.g.][]{wan2021,bonamente2006}, of the 
density of dark matter and dark energy
\citep[e.g.][]{vikhlinin2009,vikhlinin2009b,allen2004,mantz2014,mantz2022}, of the cosmological density of baryons and the associated
\emph{missing baryons} problem \citep[e.g.][]{nicastro2018,kovacs2019,spence2023}, and many more.
This  makes the
classical compounded distribution method, or fully Bayesian methods, less desirable for this class of applications. 
The model proposed in the following
aims to overcome this difficulty, by developing an approximate method of modeling systematic
errors that is both computationally tractable even with limited resources (both computational
and in terms of overall model simplicity), and more
importantly leading to a simple and accurate method of hypothesis testing.

\begin{figure}[!t]
    \centering
    \includegraphics[width=3.5in]{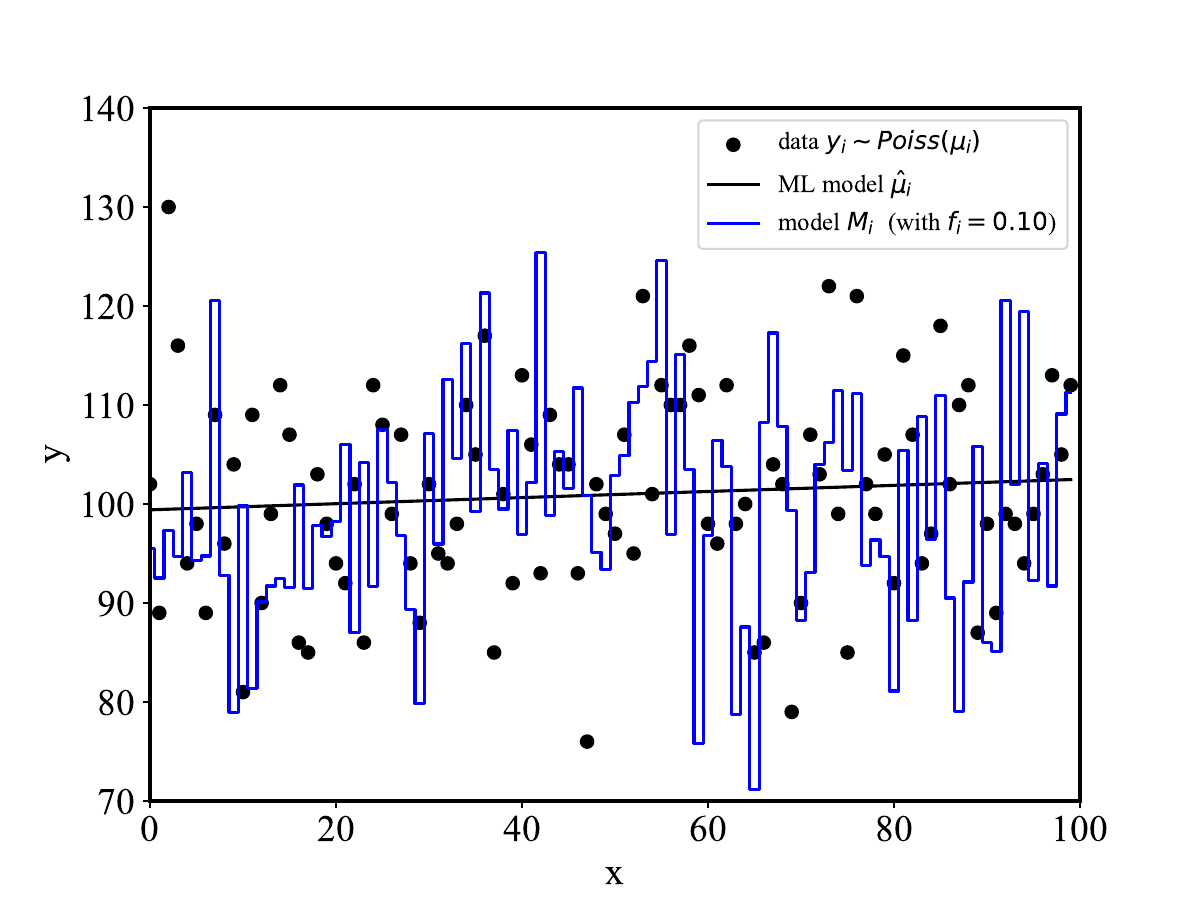}
    \caption{Illustration of the method of regression with an intrinsic model variance. Data 
    with $N=100$ with a constant parent mean $\mu=100$  (dots) were fit to a
    linear model (solid line, $\hat{\mu}_i$). The blue curve represents an instance of randomization of the
    best--fit model according to a normally--distributed $\mu_i=M_i \sim N(\hat{\mu}_i,\sigmaIntiEq^2)$ with a relative
    systematic error of $f_i \coloneqq \sigmaIntRelEq=0.1$.}
    \label{fig:dataModel}
\end{figure}

\subsection{Proposed model: the intrinsic model variance and the associated $M_i$ random variable}
\label{sec:sigmaInt}

We are now in a position to introduce our model for systematic errors
in the regression of Poisson data. 
It is proposed that, \emph{after} the usual Poisson regression is performed, 
the best--fit model $\hat{f}(x_i)$ be considered as a random variable indicated as
$M_i$, as a means to model the presence of systematic errors. 
The  mean of this random variable is naturally set to the ML estimate $\hat{\mu}_i$ itself, and 
its variance is set to a
newly introduced \emph{intrinsic model variance} parameter $\sigmaIntiEq^2$, as proposed
in \cite{bonamente2023}. 
This idea is illustrated in Fig.~\ref{fig:dataModel} for a simple linear model, whereas
the model has an intrinsic variability that is
represented by the 
blue curve, representing the post--fit randomization of the best--fit linear model according to the random variable $M_i$.

Formally, the random variable $M_i$ describes the distribution of the best--fit model $\hat{f}(x_i)$, i.e. 
\begin{equation}
 \hat{f}(x_i) \overset{d}{:=} M_i,\, \text{ with }  \E(M_i) \coloneqq \hat{\mu}_i,\;\Var(M_i)\coloneqq\sigmaIntiEq^2.
    \label{eq:Mi}
    %\label{eq:sigmaInt}
\end{equation}
Any family of distribution functions 
 with non--negative support, such as the gamma distribution, can be used,
and the
choice of distribution for $M_i$
is part of the process of modeling systematic errors. 
The choice of distribution for $M_i$ is discussed in Sec.~\ref{sec:Mi}, and it will be shown
not to be a critical task for most applications. 
Moreover, other choices could be made regarding the relationship
between the parameters of the distribution of the $M_i$ variable and
the parameters $\hat{\mu}_i$ and $\sigmaIntiEq$.
For example, the variance in Eq.~\ref{eq:Mi} could 
be further
generalized to a function of the intrinsic model variance parameter, 
i.e., $\Var(M_i)=h(\sigmaIntiEq^2)$
with $h(\cdot)$ a positive--valued function. It is believed, however, that the present
parameterization is probably sufficiently general for most applications.

The intrinsic model variance $\sigmaIntiEq^2$ is assumed to be known \emph{a priori}, e.g., from 
even limited  knowledge of
the instruments or methods used for the detection.
For example, counts measured from a detector of the type used in
high--energy astrophysics \citep[e.g.][]{lee2011} are proportional to the effective area
of the instrument, and an estimated
calibration uncertainty on the effective area may translate linearly 
to an estimate of the intrinsic model error $\sigmaIntiEq$.  Alternatively,
$\sigmaIntiEq^2$ can be  estimated from the data, as will be shown in the following.

The intrinsic model scatter
can be recast as a fraction $f_i$ of the best--fit model, whereas
\begin{equation}
    f_i \coloneqq \dfrac{\sigmaIntiEq}{\hat{\mu}_i}
    \label{eq:fi}
\end{equation}
 is the fractional amount of systematic scatter relative to the
best--fit model.
This model of systematic errors only applies to small relative systematic errors, $f_i \ll 1$
which is the common regime for a variety of applications such as the astrophysics data that motivated this model
\citep{spence2023}. In fact, large disagreements between data and model are more likely to be 
imputed to a genuine failure of the model, rather than to the presence of systematic errors.

The introduction of the random variables $M_i$ according to \eqref{eq:Mi},
and the associated intrinsic model variance $\sigma_{int,i}^2$, are the main ingredients
of this model for systematic errors in Poisson data. This model could
also be applied to other data, e.g., normally--distributed measurements, since 
definition \eqref{eq:Mi}
makes no assumptions on the distribution of the  data themselves.
Equation~\ref{eq:Mi} embodies the change of perspective in the role played by
systematic errors, i.e., the systematic variance is
associated with the \emph{model} after the regression is performed,
rather than the \emph{data} themselves.
The random variable $M_i$ plays a role that is somewhat similar to that
of a Bayesian prior on the mean (as discussed in Sec.~\ref{sec:overdispersionReview}),
in that it provides a probabilistic model that quantifies a prior belief on the intrinsic
degree of variability of the parent model.
The main difference, however, is that it is 
included in the probability model only \emph{after} the regression is performed,
and it does not have an effect on it.
Therefore the random variable $M_i$ is used primarily for its effect on the
distribution of the null--hypothesis goodness--of--fit statistic, 
as will be explained
in detail in Sec.~\ref{sec:cmin}.

\section{The goodness--of--fit statistic for the Poisson regression with systematic errors}
\label{sec:cmin}
\def\yMi{\tilde{y}_i}

{The introduction of the $M_i$ random variable to model systematic errors is now
incorporated into a statistical model for the regression of Poisson data. The main aim of this model is to
derive a \gof\ likelihood--ratio statistic that generalizes the usual \cmin, when systematic
errors are present.
}
\subsection{The quasi \ml\ estimation}
\label{sec:quasiMLE}
The choice of introducing an intrinsic model variance and the associated 
random variables $M_i$ to describe the model 
after the ML regression is performed (see Sec.~\ref{sec:sigmaInt}) is  borne out of convenience.
With that assumption 
the data $y_i$ can in fact be viewed as 
being distributed according to the compounded distribution
\begin{equation}
    %\yMi 
    y_i \sim \Poiss(M_i)
    \label{eq:yCompoundData}
\end{equation}
with $M_i$ given by  \eqref{eq:Mi}, in place of the usual $y_i \sim \Poiss(\mu_i)$.
Specifically, the moments requirements for $M_i$  imply 
that the expectation remains $\E(y_i)=\mu_i$ under $H_0$, while $\Var(y_i) > \mu_i$ according to the usual
compounding formulas.
This means that the data 
can be viewed as retaining the same parent means, but with
larger variance (i.e., overdispersed), 
as was the intent of the method (see Sec.~\ref{sec:sysErrDefinition}).

With this interpretation, which will be further justified is Sec.~\ref{sec:gof} below,
the usual Poisson \ml\ method of estimation that we use for the regression becomes
a \emph{quasi maximum--likelihood} estimation \citep[e.g., Sec.~3.2.2 of][]{cameron2013},
which is guaranteed to continue providing consistent estimates
assuming that the means are correct according to the null hypothesis \citep[as shown in, e.g.][]{gourieroux1984b}.
In other words, the regression can be
performed by minimizing the usual \DP\ statistic according to \eqref{eq:DP}, and therefore
there is no additional burden to the task of identifying the best--fit model in the presence of systematic errors.

Minimization of the \DP\ statistic must be achieved with numerical methods, even for
the simple linear model \citep[e.g.][]{cameron2013,bonamente2022book}. 
In fact, there is no fully analytical solution to the \ml\ equations
beyond the simple one--parameter constant model, when the parameter is trivially estimated by the sample
mean.  A semi--analytical solution to the \ml\
fit for the two--parameter linear model for Poisson data was provided in \cite{bonamente2022}, including analytical estimates
of the covariance matrix \citep{bonamente2023b}, which can reduce the computational
burden for the linear model. For more complex models, the Poisson
log--likelihood must be maximized via numerical methods.

\subsection{ The \gof\ statistic}
\label{sec:gof}
We now turn to the  
goodness--of--fit statistic
in the presence of systematic errors, which we refer to as \cminsys.
Given that the $M_i$ variables replace the simple
$\hat{\mu}_i$ estimates in this model of systematic errors,
this new \gof\ statistic is formally $ \cminsyseq \coloneqq \cmineq(M_i)$ according to \eqref{eq:cmin}, i.e.,
\begin{equation}
\begin{aligned}
    \cminsyseq = %&\cmineq(M_i)  \coloneqq \\
    &
     2 \sum_{i=1}^N \left( y_i \ln \left(\dfrac{y_i}{M_i}\right) - (y_i-M_i) \right).
    \end{aligned}
  \label{eq:cminMi}
\end{equation}
This statistic is therefore also immediately obtained as a likelihood--ratio
statistic from the joint log--likelihood of the data distributed as 
in \eqref{eq:yCompoundData}, thus supporting our interpretation of Sec.~\ref{sec:quasiMLE}.

It is also immediate to see that \eqref{eq:cminMi} can be written as the \emph{sum} of two terms.
Referring to the statistic as  $Z$ for convenience, we can set
\begin{equation}
    \cminsyseq \coloneqq Z=X+Y,
    \label{eq:CminMi}
\end{equation}
%$Z$ is  the sum of two random variables, 
of which $X \coloneqq \cmineq(\hat{\mu}_i)$ represents the usual $D_P$
statistic according to \eqref{eq:DP}. The additional
statistic $Y$  is associated with the the random variable $M_i$,
and it is defined by
\begin{equation}
\begin{aligned}
    Y = & \cmineq(M_i)-\cmineq(\hat{\mu}_{i}) = \\
    & 2 \sum_{i=1}^N (M_i-\hat{\mu}_i)-y_i \ln \left(\dfrac{M_i}{\hat{\mu}_i}\right).
    \end{aligned}
    \label{eq:YStat}
\end{equation}
It is clear that, when there
are no systematic errors, $Y$ vanishes as $M_i=\hat{\mu}_i$ identically.
Moreover, since $M_i$  is independent of
the data $y_i$, $Y$ is asymptotically normal in the extensive data limit (i.e., large $N$) by the central limit theorem. ~\footnote{This is true provided each term of $Y$ has finite
mean and variance according to the choice of distribution in \eqref{eq:Mi}, which is
usually the case for most distributions.} 
Accordingly, we can set the asymptotic distribution of
$Y$ as
\begin{equation}
    Y \overset{a}{\sim} N({\mu}_C,{\sigma}^2_{C}),
    \label{eq:YDistr}
\end{equation}
where the two parameters
of the distribution represent the additional mean and  variance 
that are induced by the presence of systematic errors. 

{ This model of systematic errors has therefore achieved our primary goal to
provide a simple generalization for the \gof\ statistic for the Poisson regression
in the presence of systematic errors.}
The first
contribution to the new \gof\ statistic $Z=\cminsyseq$ is thus the usual Poisson log--likelihood $X=D_P$,
whose distribution is independent of the shape of the model
$\mu_i=f(x_i; \theta)$ according to the Wilks theorem. Therefore, the exact parameterization
does not play a significant role in the determination of the overall distribution for \cminsys,
provided that basic requirements on model identifiability are satisfied 
\citep[e.g.][]{li2024, hoadley1971, dale1986}.

The additional contribution $Y$ has a simple normal asymptotic distribution and, moreover, it is \emph{independent} of $X$ under the null hypothesis. This is the case for the following
reasons. First, as we just remarked, the Wilks theorem assures that 
the null--hypothesis distribution of $X$ is
not dependent on $\hat{\mu}_i$. Second, $Y$ is by construction not dependent on the data $y_i\sim \Poiss(\mu_i)$, since the $M_i$ random variable defined in \eqref{eq:Mi}
depends only on the best-fit means $\hat{\mu}_i$, and on an intrinsic model variance that
is assumed to be known a priori.

\subsection{Estimation of the parameters of $Y$}
In order to estimate the asymptotic
distribution of the Poisson \gof\ statistic in the presence of systematic errors, as given in \eqref{eq:cminMi},
it is therefore necessary to simply estimate the mean and variance of $Y$ in \eqref{eq:YDistr}.  Detailed calculations to accomplish this task are provided in
Appendix~\ref{app:YMoments}, and key results are summarized in the following.

The moments of the $Y$ distribution can be obtained via a Taylor series expansion of the
logarithmic terms in \eqref{eq:YStat}, 
\begin{equation}
\begin{aligned}
        Y 
        %& =
        %2 \sum_{i=1}^N (M_i-\hat{\mu}_i)-y_i \ln \left(\dfrac{M_i}{\hat{\mu}_i}\right) \\
        %&  
        \simeq 2 \sum_{i=1}^N (M_i-\hat{\mu}_i) \left(1- \dfrac{y_i}{\hat{\mu}_i}\right)
        + \sum_{i=1}^N y_i\left(\dfrac{M_i-\hat{\mu}_i}{\hat{\mu}_i}\right)^2, 
        %+ o\left(\left(\dfrac{M_i-\hat{\mu}_i}{\hat{\mu}_i}\right)^2\right),
\end{aligned}
\label{eq:YApprox}
\end{equation}
by retaining second--order terms in \eqref{eq:DP}. 
Given that $\E(M_i)= \hat{\mu}_i$, the expectation of the
terms in the first sum of \eqref{eq:YApprox} are null, leaving a second--order correction of 
\begin{equation}
    \E(Y) = \mu_C \simeq \sum_{i=1}^N  \mu_i\, f_i^2,
    \label{eq:muC}
\end{equation}
where $\Var(M_i)=\sigmaIntiEq^2$ according to \eqref{eq:Mi}. 
The ratio $f_i= \sigmaIntiEq/ \hat{\mu}_i$ is the fractional amount of scatter, and it is intended to be a  number that is much smaller than one.
This expectation will be referred to as the
\emph{bias parameter} $\mu_C$ associated with \cminsys\ statistic.
This expectation can therefore be estimated
from the data via
\begin{equation}
    \hat{\mu}_C \simeq \sum_{i=1}^N  y_i\, f_i^2.
    \label{eq:muHatC}
\end{equation}

The variance of $Y$
is referred to as the 
\emph{overdispersion parameter} $\sigma^2_{C}$, and it evaluates to
\begin{equation}
\begin{aligned} 
  \Var(Y) &= \sigma^2_{C} \simeq  4 \sum_{i=1}^N \mu_i\,  f_i^2 \\ &  
  + \sum_{i=1}^N (\mu_i^2+\mu_i)\, \kurt(M_i)\, f_i^4 
  - \sum_{i=1}^N \mu_i^2\, f_i ^4,
  \end{aligned}
  \label{eq:sigmaC}
\end{equation}
with $\kurt(M_i)$ the kurtosis (or normalized central moment of the fourth order) of $M_i$, e.g.,
$\kurt(M_i)=3$ for a normal distribution.
The second and third term in the right--hand side are the second--order correction to the 
approximate results
already provided in \cite{bonamente2023}.
This approximation applies again for small
relative errors, $f_i \ll 1$.
This variance can be estimated from the data according to 
\begin{equation}
   \hat{\sigma}^2_{C} \simeq  4 \sum_{i=1}^N y_i\, f_i^2 
   + \sum_{i=1}^N y_i^2\,  f_i^4 \cdot \left(\kurt(M_i) -1\right),
  %- \sum_{i=1}^N y_i^2\, f_i^4,
  \label{eq:sigmaHatC}
\end{equation}
see Appendix~\ref{app:YMoments} for details.

Equations~\ref{eq:muHatC} and \ref{eq:sigmaHatC} are the main results for the
estimation of the $Y$ distribution, and they relate the key parameter of the model for systematic errors
(i.e., the intrinsic model variance) to the bias and overdispersion parameters for the \cminsys\
statistic in the presence of systematic errors. In principle, one may retain higher--order terms
in the power series expansions. The accuracy of these approximations is tested in Sec.~\ref{sec:cminSym},
to show that for typical values of systematic uncertainties, the second--order corrections presented in this
paper may be sufficient for most applications.

\subsection{Asymptotic distributions}
\label{sec:cminDistr}
In general, a simple closed form for the asymptotic distribution of \cminsys\ cannot be given in all regimes,
as discussed in \cite{li2024}. However, useful approximations can be given that cover most practical
data analysis cases. 

In the large--mean regime, $X \approx \chi^2(\nu)$ due to the approximation $\Poiss(\mu) \approx N(\mu, 2 \mu)$ for the
 distribution of the data
 (this is true even for dataset with small $N$).
 Provided that $Y$ is approximately normally distributed for that value of $N$, then
 $Y \approx N(\hat{\mu}_C,\hat{\sigma}^2_C)$,  and it follows that in the large--mean
 regime  \cminsys\ is approximately distributed like the sum of a
 chi--squared and an independent normal variable. The resulting
 distribution is referred to as the \emph{overdispersed chi--squared distribution} $B$, 
 \begin{equation}
     \cminsyseq \sim B(\nu,\hat{\mu}_C,\hat{\sigma}^2_C),
 \end{equation}
 which is a special case of the gamma--normal family of distributions,
 with a probability distribution function that is the convolution of
 independent $\chi^2(\nu)$ and $N(\hat{\mu}_C,\hat{\sigma}^2_C)$  distributions \citep[see][for properties of this distribution] {bonamente2024properties}.

For applications with extensive data
and in the large--mean regime,
it is possible to further approximate $X \overset{a}{\sim} N(\nu, 2\nu)$,
therefore bypassing the use of the overdispersed chi--squared distribution. Therefore \cminsys\ is asymptotically normal with mean and variance equal to the sum of means and variances of the two 
contributing statistics,~\footnote{This can be  immediately seen from the property of
infinite divisibility of the normal distribution, or that the convolution of two normal distributions
remains normal.}
\begin{equation}
\cminsyseq \overset{a}{\sim} N(\nu+\hat{\mu}_C, 2 \nu+ \hat{\sigma}^2_C).    
\label{eq:ZDistAsymptotic}
\end{equation}
Therefore, for most applications
in the large--mean and extensive data regime, this model of systematic errors
for Poisson data results in a straightforward extension of the $\chi^2(\nu)$ distribution
for the \cmin\ fit statistic in that regime, with the simple use of \eqref{eq:ZDistAsymptotic} as the 
null--hypothesis distribution for the \gof\ statistic.

In the low--mean regime, however, additional considerations are needed
even in the asymptotic limit of extensive data.
\cite{li2024} has shown that the Poisson deviance $X=D_P$ remains asymptotically normal, 
but with mean and variance that will differ significantly from the simple large--mean 
values assumed in \eqref{eq:ZDistAsymptotic}. For extensive data in the low--mean regime,
\begin{equation}
    \cminsyseq \overset{a}{\sim}  N(\E(\cmineq|\hat{\theta})+\hat{\mu}_C, \Var(\cmineq|\hat{\theta})+ \hat{\sigma}^2_C),
    \label{eq:ZDistAsymptoticLi}
\end{equation}
where the conditional mean $\E(\cmineq|\hat{\theta})$ and 
the conditional variance $\Var(\cmineq|\hat{\theta})$  of \cmin\ are provided by Theorems~{8--10} of \cite{li2024}.
In the low--mean regime, there is therefore only the additional burden of evaluating the
appropriate moments for \cmin, and the asymptotic distribution of \cminsys\ remains normal
for extensive data.

\subsection{ Distribution for $M_i$ and other considerations}
\label{sec:Mi}
Provided that the relative level of systematic errors is small, $f_i \ll1$ as assumed throughout, the choice
of distribution for $M_i$ is not critical.   
According to approximation
\eqref{eq:YApprox}, $\E(Y)$ depends only on the
 first and second order moments for $M_i$ 
and it is independent of other moments of the distribution
of $M_i$, and \eqref{eq:Mi} ensures that $\E(Y)$ is  independent of the choice for the distribution of $M_i$. Moreover, $\Var(Y)$ has only a mild dependence on higher order moments, according to \eqref{eq:sigmaHatC} or \eqref{eq:VarYApp}.
It is nonetheless useful to illustrate in detail how the choice of distribution affects the estimate of
the variance of $Y$. For this purpose, we use a gamma--distributed $M_i$ variable.

Let $M_i$ be distributed according to a gamma distribution with rate parameter $\alpha$
and shape parameter $r$, indicated by $M_i \sim {\rm gamma}(\alpha,r)$. If the variable is required to have
an expectation $\E(M_i)\coloneqq r/\alpha=\hat{\mu_i}$ and a variance $\Var(M_i) \coloneqq r/\alpha^2=\sigmaIntiEq^2$ according to \eqref{eq:Mi}, 
then the two parameters of the distribution are respectively  $\alpha={\hat{\mu}_i}/{\sigmaIntiEq^2}$
and $r={\hat{\mu}_i^2}/{\sigmaIntiEq^2}$.
The variance $\hat{\sigma}^2_C$ according to
\eqref{eq:sigmaHatC}  then becomes
\begin{equation}
  %\Var(Y) = 
  \hat{\sigma}^2_{C} \simeq  4 \sum_{i=1}^N y_i\, f_i^2 
  +\left(3+6 f_i^2 -1\right) \cdot  \sum_{i=1}^N y_i^2\, f_i^4,
  \label{eq:sigmaHatCgamma}
\end{equation}
where the term $6 f_i^2$
which  reflects the excess kurtosis
of the gamma distribution  has a small effect, given that $f_i \ll 1$.
Any other choice for the distribution of $M_i$ can be treated in a similar way, and the
resulting estimates for the variance of the $Y$ distribution is in general provided 
by \eqref{eq:VarYApp}. 

Alternatively, a normal distribution for $M_i$ would reflect the assumption that there may be multiple sources for this variance
that are likely to yield a normal distribution, according to the central limit theorem.
However, the fact that a normal distribution allows negative values makes this choice
 reasonable only in the limit of $\sigmaIntRelEq\ll1$,
whereas one ignores negative values of the normal distribution that are untenable for
the mean of a  Poisson distribution. 

Finally, it may be useful to remark that the normal approximation for the $Y$ statistic is simply a result
of the central limit theorem for extensive data. This conclusion applies regardless of choice for
distribution of $M_i$, provided the distribution has finite mean and variance, as is the
case for most distributions including the normal and gamma.

\section{Monte Carlo tests} 
\label{sec:cminSym}

The statistical model of systematic errors presented in Sects.~\ref{sec:model} and \ref{sec:cmin} is now tested with the aid of Monte Carlo numerical simulations that are representative of real data analysis cases, with the goal of illustrating the overall accuracy of this model to predict the distribution of \cminsys, and its limitations.
Given the independence of the $X=\cmineq(\hat{\mu}_i)$ statistic on the parameterization of the regression 
model $f(x)$, as 
discussed in Sec.~\ref{sec:cmin}, it is sufficient to test the results with simple analytic
models. For this paper, we choose a constant model with $m=1$ parameter, and a linear model
with $m=2$ parameters.
The Monte Carlo simulations are aimed at validating the distributions for the fit
statistics $X$, $Y$ and $Z$ with these simple models.

\subsection{Linear model with 10\%  and 5\% systematic errors}
\label{sec:Sim1}
For these simulations, a  dataset with $N=100$ independent data points is drawn from a constant model with a parent mean $\mu=100$, in such a way that the Poisson data are in the extensive and large--mean regime. The data are fit to the same constant model, and also to a linear model $y=a+b\,x$  to obtain the \ml\ estimates $\hat{\mu}_i$. 

First, the $X$ statistic
is calculated according to \eqref{eq:DP}. 
Next, in order to simulate the presence of systematic errors,
the best--fit models (both full and reduced) are \emph{randomized} at each point according to a normal distribution for $M_i$ with parameters \eqref{eq:Mi}, using a fixed value of $f_i=0.1$ representing a 10\% level of systematic errors.  
The fit statistic with these randomized models
are referred to as $Z$ and $Z_r$, with "r" denoting the fit to the reduced model. The contribution to the fit statistic due to the systematic error is then $Y=Z-X$. The same procedure was repeated 500 times to obtain the empirical cumulative distribution function (eCDF) of the statistics $Z=\cminsyseq$, $X$ and $Y$. 

Tests of the agreement between the empirical and predicted distributions of the $X$, $Y$ and $Z$ statistics were performed with the Kolmogorov--Smirnov one--sample statistic $D_N=\max(D^-,D^+)$ \citep{kolmogorov1933}. 
The fit statistics $X$ and $X_r$ in the absence of systematic errors 
are expected to be distributed as $\chi^2(N-m)$ according to the Wilks theorem.  In this limit, this distribution is well approximated by a normal distribution of same mean and variance.
In the same limit, the $Y$ and $Y_r$ statistics are approximated
by normal distributions with mean and variance according to \eqref{eq:muHatC} and \eqref{eq:sigmaHatC}, and the $Z$ and $Z_r$ statistics also by normal distributions according to \eqref{eq:ZDistAsymptotic}. For all statistics $X$, $Y$ and $Z$,
the corresponding $D_N$ statistics for the comparison between the eCDF and the expected distribution have large $p$--values, indicating good agreement.

 Additional Monte Carlo simulations similar to the one reported above 
 were repeated with the level of systematic errors 
now set at 5\%, or $f_i = 0.05$. As for the previous case, we obtained good agreement between theoretical and simulated statistics, according to the respective $D_N$ statistics.
% Numers to be updated with these: DONE
%Experimental: Y: 25.188 +- 11.124, Yr: 25.183 +- 11.223
%Expected: Y: 25.000 +- 10.897
%Bias and overdispersion in Cmin with randomization 25.188201346931333 11.124008280007084
%Summary of KS tests for Cmin (full)
%X: 0.030 (0.7439); Y: 0.026 (0.8716); Cmin=X+Y:  0.034 (0.5876)
%Summary of KS tests for Cmin (reduced)
%X: 0.029 (0.7688); Y: 0.029 (0.7886); Cmin=X+Y:  0.035 (0.5524)
The  case of 5\% systematic errors 
better represents a typical real--life data analysis situation 
where systematic errors may be invoked, rather than the case of 10\% systematics. 
In fact, for  $N=100$ and $\mu=100$, a 5\% systematic error %($f_i=0.05$) 
leads to a 
bias $\hat{\mu}_C=25$ that is at the level of 25\% of the expected 
null--hypothesis \cmin\ statistic (with $\E(X)=N-m \simeq 100$).
This figure rises to $\sim$100\% for the 10\% systematic errors, i.e., a bias of $\hat{\mu}_C=100$.

\subsection{Linear model with GLM log--link}
\label{sec:GLM}

Generalized linear models are commonly used in a variety of disciplines
such as econometrics \citep[e.g.][]{cameron2013}, and occasionally also in astronomy \citep[e.g.][]{desouza2015}.
In GLM models,  the usual linear model with the two parameters $\theta=(a,b)$ is used witha linear predictor $\eta_i=a+b\,x_i$ and with Poisson mean $\mu_i=\exp(\eta_i)$, which
corresponds to the canonical log link in the GLM notation, $\ln \mu_i = \eta_i$. This is a common
choice for the analysis of Poisson data, to avoid possible non--positive means with the usual linear model of Sec.~\ref{sec:Sim1}.
We therefore performed simulations with the same parameters as for the usual linear model
that was used in the previous section (with $N=100$ and $\mu=100$), and obtain statistically indistinguishable results
from the results with the traditional linear model.
Quantitative results with  both the traditional linear model and the log--link model are presented in Sec.~\ref{sec:SimY} below.

A GLM model with the canonical logarithmic link corresponds to a Pareto or power--law
model with Poisson mean $\mu_i=A\cdot x_{i,r} ^b$, with $A=\exp(a)$ and when the independent
variable is on an exponential scale, $x_{i,r}=\exp(x_i)$.
It is necessary to highlight that, for any physically--based model $y=f(x)$, the usual
linear model and the GLM linear model with a canonical log--link are \emph{not} equivalent. 
Given that the linear model with log--link is equivalent to a power--law model with
exponentially--spaced independent variables $x_{i,r}$, the GLM log--link linear model  provides additional evidence of
the independence of the results from model parameterization.

\subsection{Comprehensive tests  of the $Y$ statistic}
\label{sec:SimY}
Comprehensive Monte Carlo tests for the mean and variance of the distribution
of $Y$ were performed as follows. A selected number of representative 
values for the $N$, $\mu$ and $f$
parameters were chosen, and 500 Monte Carlo realizations of the a constant model
were used to generate random datasets that were then fitted to the same constant model, and also to a two--parameter linear model, same as for the 
simulations presented in Sec.~\ref{sec:Sim1}.

\begin{figure*}
    \centering
    \includegraphics[width=3.5in]{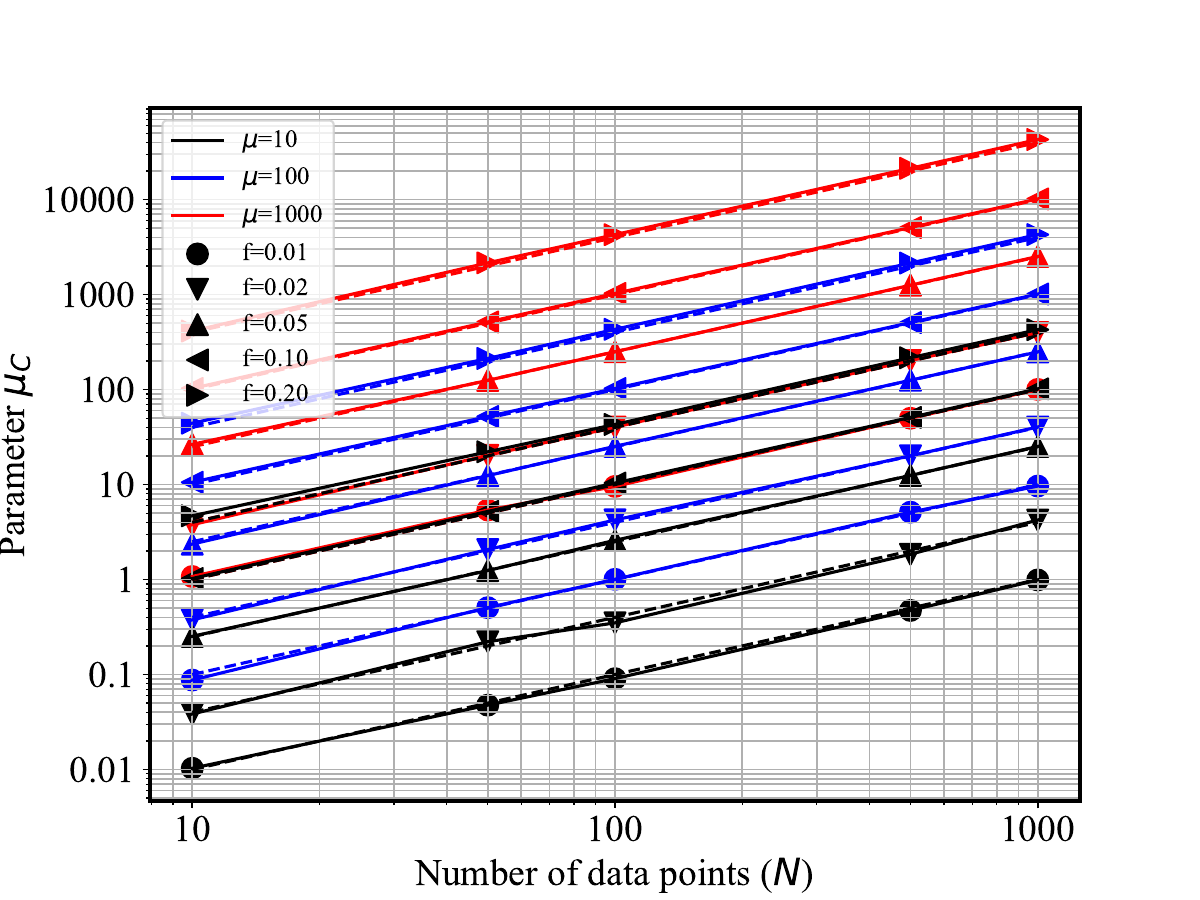}
    \includegraphics[width=3.5in]{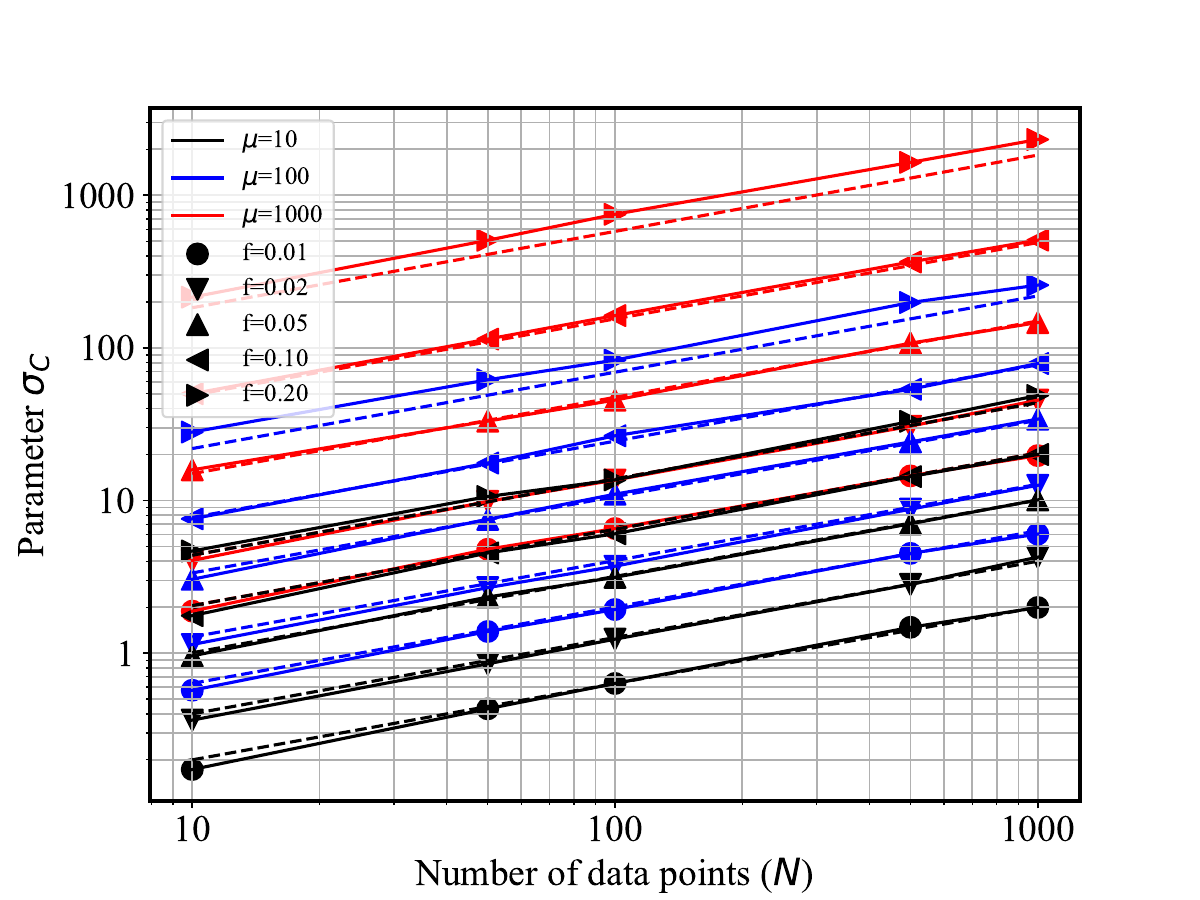}
    \caption{Results of  the Monte Carlo simulations of the $Y$ variable for a dataset with 
    a constant parent model. The parent value of the Poisson mean are coded according to
    color, and the symbols (also color--coded) are used to distinguish the values $f=0.01, 0.02, 0.05, 0.10$ and 0.20
    for the fractional systematic error. Dashed lines are the
theoretical curves with the approximations of \eqref{eq:muHatC} and \eqref{eq:sigmaHatC}. The $f=0.20$ curves with color--coded
    symbols $\blacktriangleright$ are used to illustrate
    the departure from the approximations when the systematic errors are too large, and are not used in applications. Datasets with $N=10$, 50, 100, 500 and 1000 independent datapoints were simulated.}
    \label{fig:MC}
\end{figure*}

The mean
and the variance of the resulting $Y$ statistic for the fit to the linear model are
shown in Fig.~\ref{fig:MC}, along with the predictions from \eqref{eq:muHatC} and \eqref{eq:sigmaHatC} that are reported as dashed lines. 
For all parameters, the Monte Carlo estimates of $\hat{\mu}_C$ and $\hat{\sigma}_C$  agree typically with the
model predictions to within $\pm 10$\% for $f_i \leq 0.1$, and often within just a few percent. 
This degree of agreement indicates that the $Y$ random variable is
sufficiently accurately described by \eqref{eq:YDistr}, in a wide range of model parameters that are typical
for data analysis applications. Identical results
were found for the $Y_r$ statistic for the fit to the reduced model.

The range of parameters for the simulation was limited by design to $f_i \leq 0.1$, plus the case of $f_i=0.2$ that is used 
to illustrate the errors in the estimates for larger values of systematic errors.
Table~\ref{tab:MCsummary} reports the average fractional deviations $\eta_{\mu}=\hat{\mu}_C/\mu_C-1$ and 
$\eta_{\sigma}=\hat{\sigma}_C/\sigma_C-1$, and their standard deviations, according to the results shown in Fig.~\ref{fig:MC}.
A 20\% systematic error is outside the domain of approximations presented in this
paper, and therefore the deviations seen in Fig.~\ref{fig:MC} and Table~\ref{tab:MCsummary} for $f_i=0.2$ is simply a result of the inaccuracy of the Taylor series approximation, see App.~\eqref{eq:YApp}. In fact, large values of this parameter
are not of practical use in data analysis applications, as already remarked in  Sec.~\ref{sec:Sim1}.
Also, the low--mean regime $\mu \leq 10$ is not tested by these simulations. 

In addition, another set of Monte Carlo simulations similar to those illustrated in Fig.~\ref{fig:MC}
was performed using the linear GLM with canonical link introduced in Sec.~\ref{sec:GLM}.
In this case,  we used
a logarithmically--spaced set of independent variables
$x_i$ between 0 and $\log(N)$, to avoid possible overflow issues when implementing the exponential mean.
The results of those simulations are statistically indistinguishable from those with the usual
linear model,  see Table~\ref{tab:MCsummary}.

\begin{table}[]
    \caption{Fractional differences $\eta_{\mu}$ and $\eta_{\sigma}$ between the Monte Carlo simulations of $\mu_C$ and $\sigma_C$ (Fig.~\ref{fig:MC}) and their
    predictions according to \eqref{eq:muHatC} and \eqref{eq:sigmaHatC}.}
    \label{tab:MCsummary}
    \centering
    \begin{tabular}{c|cc}
    \hline
    \hline
    $f=\sigmaIntRelEq$ & $\eta_{\mu} $ & $\eta_{\sigma}$ \\
    \hline
    & \multicolumn{2}{c}{Linear model}\\
    \hline
        0.01 &  -0.012$\pm$-0.055 & -0.055$\pm$0.055\\
        0.02 & -0.004$\pm$0.058  & -0.030$\pm$0.046\\
        0.05 &  0.004$\pm$0.023 & -0.003$\pm$0.037\\
        0.10 & 0.027$\pm$0.012 & 0.002$\pm$0.053\\
        \hline
        0.20 & 0.081$\pm$0.029 & 0.184$\pm$0.095\\
        \hline
        \hline
        & \multicolumn{2}{c}{GLM with log--link}\\
        \hline
        0.01 &  0.022$\pm$0.160 & -0.013$\pm$-0.047\\
        0.02 & 0.021$\pm$0.083  & -0.024$\pm$0.057\\
        0.05 &  0.010$\pm$0.041 & 0.000$\pm$0.044\\
        0.10 & 0.010$\pm$0.030 & 0.002$\pm$0.049\\
        \hline
        0.20 & 0.068$\pm$0.015 & 0.148$\pm$0.082\\
         \hline
         \hline
    \end{tabular}
\end{table}

%index=10
%sys=0.01: deviation -0.012+-0.055
%sys=0.02: deviation -0.004+-0.058
%sys=0.05: deviation 0.004+-0.023
%sys=0.10: deviation 0.027+-0.012
%sys=0.20: deviation 0.081+-0.029
%index=12
%sys=0.01: deviation -0.055+--0.055
%sys=0.02: deviation -0.030+-0.046
%sys=0.05: deviation -0.003+-0.037
%sys=0.10: deviation 0.002+-0.053
%sys=0.20: deviation 0.184+-0.095

The agreement between the Monte Carlo simulations for systematic errors  with
$f_i \leq 0.1$ and the predictions according to the theory
presented in Sec.~\ref{sec:sigmaInt} and in Sec.~\ref{sec:cmin} validates this model
 as an accurate tool to estimate the goodness of fit for the 
regression of Poisson data in the presence of systematic errors. 

\section{Methods of hypothesis testing}
\label{sec:applications}

In this section we provide a summary of the use of this model
for two classes of applications. 
In the first case, a known value for the intrinsic model variance \sigmaInt\ can be used for the usual hypothesis testing of a measured value of \cmin. In the second case, when it is not possible or
convenient to estimate the intrinsic variance a priori, the distribution of \cmin\ presented in this paper
can be used to estimate the intrinsic variance itself, under the assumption that the null hypothesis is correct. We then provide an application to real--life astronomical data.

\subsection{Hypothesis testing with known systematic errors}
\label{sec:hypothesisTesting}
In many data analysis situations, it is possible to estimate \emph{a priori}
the level of systematic errors present in the measurements. {This is the regime that was 
envisioned for the
model of systematic errors developed in Sec.~\ref{sec:model}, where the $\sigmaIntiEq$ parameters
are assumed to be known.}
For example, inherent uncertainties associated with
 either or both the photon--counting device and the parent astrophysical process that generates the photons are common for the detection
of high--energy photons from distant quasars, such as in the recent applications by \cite{spence2023,nicastro2018,kovacs2019}, among many others. In that class of applications,
it is often common to use a simple power--law 
or other phenomenological models such as a spline, 
to model the
number of photons detected as a function of wavelength. The parent model is likely to be more complex and often unknown, giving rise to the need for a source of 
systematic error in the Poisson regression, as opposed to the employment of
an alternative model. 

For example, in \cite{spence2023} we estimated a level of systematic errors in the calibration of the so--called effective area of
the CCD detector at the few \% level for the main instrument used for
those data. Given that the effective area of the instrument is
linearly related to the measured counts, this uncertainty could be used as an a priori value of
the parameter, say $f_i \simeq 0.02$, to test whether the resulting 
fit statistic \cmin\ is consistent with that value of the intrinsic model variance.

In this class of applications, assuming  extensive data, 
hypothesis testing consists of two steps: \\
(1) First, 
the bias and overdispersion parameters $\mu_C$ and $\sigma^2_C$ of the \cminsys\ statistic are estimated from the data according to \eqref{eq:muHatC} and \eqref{eq:sigmaHatC},
for a given value of the chosen level of systematic errors $f_i$;\\
(2) Then,
the relevant parent distribution for $\cminsyseq$, as discussed in Sec.~\ref{sec:cminDistr},
is used to perform the usual one--sided null--hypothesis test with the measured value of \cmin, for a given
$p$--value. 

\subsection{Estimate of the intrinsic model variance from data}
\label{sec:estimateF}
In certain experimental cases, it is not possible or interesting to provide an accurate estimate
of the level of systematic errors present in the problem {a priori}. This situation arises
when the data regression yields an unacceptably high value of $\cmineq$, 
and the data analyst (a) may not provide an alternative model for the regression, or (b) is interested in
evaluating the level of systematic errors relative to such model, following the belief that the data
are in fact accurately described by the model at hand. 

In this class of applications, the parent distribution of \cminsys\ according to \eqref{eq:ZDistAsymptotic}, or
\eqref{eq:ZDistAsymptoticLi} in the low--mean regime,
can be used to estimate the intrinsic model variance $\sigmaIntiEq^2$, i.e., the level of systematic errors, \emph{assuming} that the regression model {is} the parent model. 
In this section, 
the measured \gof\ statistic will be referred to as lower--case \cminVal, to distinguish it from the
statistic itself.
We also assume that the intrinsic level of systematic errors $f_i$ is
constant among the data points, and we indicate this
value as $f$.

A point estimate for the average level of systematic errors $f$
can be obtained by requiring 
$\E(Z)=\cminValEq$, i.e., the measurement equals the expectation of the parent distribution. 
In general, the values of the mean and 
variance of $X$ were discussed in Sec.~\ref{sec:cminDistr}, and in the large--mean regime 
they are $\E(X)=(N-m)$ and $\Var(X)=2(N-m)$.
It follows that the bias parameter can be estimated as
\begin{equation}
\hat{\mu}_C = \cminValEq-\E(X) >0.
\label{eq:muCEst}
\end{equation}
Then, using \eqref{eq:muHatC} and assuming a constant value for $f$, 
the estimate is simply obtained as
\begin{equation}
\widehat{f\,}=
\sqrt{\dfrac{\hat{\mu}_C}{\sum_{i=1}^N y_i }}.
\label{eq:sigmaIntEst}
    %\E_i(Y) = \hat{\mu}_C \simeq \sum_{i=1}^N  y_i \left(\dfrac{\sigma_{i,int} }{\hat{\mu}_i} \right)^2.
\end{equation}
For example, a value of \cminVal=125  
for a dataset with $N=100$ and a parent mean
$\mu=100$, leads to an estimated $\hat{f}=0.052$ according to \eqref{eq:sigmaIntEst}, similar to the one assumed for the Monte Carlo simulations with $f_i=0.05$. 

Confidence intervals for $f$ can be obtained by requiring that
the measurement \cminVal\ corresponds to specified quantiles of the parent 
distribution for $Z$. For example, for a 68\% confidence interval, the condition
\begin{equation}
\cminValEq=\E(X) + \hat{\mu}_C \pm \sqrt{\Var(X)+\hat{\sigma}^2_C}
\label{eq:sigmaIntConf1}
\end{equation}
can be used to obtain the lower and upper limit of the confidence interval
for    \sigmaIntRel, upon which the bias and overdispersion parameters in
\eqref{eq:sigmaIntConf1} depend. 
The equation is quadratic in $f$,
and it can be further simplified if the overdispersion parameter $\hat{\sigma}^2_{C}$ is held fixed at the 
point estimate \eqref{eq:sigmaHatC}. With this approximation, 
confidence interval{\color{blue}s} on the $f$ parameter are
estimated via
\begin{equation}
  f^2_{\text{up,lo}}  \simeq \dfrac{\cminValEq-\E(X) \pm \alpha \times \sqrt{\Var(X) + \hat{\sigma}^2_{C}}}{\sum_{i=1}^N y_i },
    \label{eq:sigmaIntConf2}
\end{equation}
i.e., the lower and upper bounds of the confidence intervals are obtained from \eqref{eq:sigmaIntConf2}, with $\alpha=1$ for a 68\% confidence level.
Confidence intervals at a different level of probability $p$
(i.e, $p=0.9$ or 0.99) can be obtained by multiplying the last term in the numerator
of \eqref{eq:sigmaIntConf2} by the proper factor $\alpha$ corresponding to the 
$(1+p)/2$ quantile of an $N(0,1)$ distribution (e.g., $\alpha=1.65$ and 2.58 for $p=0.9$ and 0.99, respectively). Continuing with the same example as above, the 68\% confidence interval according to \eqref{eq:sigmaIntConf2} is given by $f=(0.031,0.067)$, and the estimate of the systematic error would be reported as
$\hat{f}=0.052^{+0.15}_{-0.21}$. It is clear that \eqref{eq:sigmaIntConf2} may yield a negative number for the lower limit; that is 
an indication that
no systematic errors are
required in that case (i.e., a lower confidence value of zero).

{In principle, it may be possible to estimate the unknown $\sigmaIntiEq$ parameters following
a direct \ml\ method, which depends on the choice of the compounding distribution for the random
variable $M_i$ (see Sec.~\ref{sec:Mi}). For example, a compounding gamma distribution results in a negative binomial distribution for the data, as discussed in Sec.~\ref{sec:sigmaInt}, and one could proceed with a \ml\ method to estimate
the overdispersion parameter of the gamma distribution \citep[e.g., following the methods discussed in ][]{fisher1941, piegorsch1990, wang2010}. Other methods are also
are available to estimate the compounding distribution of a Poisson variable \citep[e.g.][]{tucker1963, simar1976}. The approximate method to estimate the overdispersion parameter presented in this section, on the other hand,
has the advantage of being distribution--independent,
provided conditions \eqref{eq:Mi} on the choice of $M_i$ are met. This method
can therefore be used immediately for virtually
any application, without the additional complications associated with the use of a direct \ml\ method for this task.}

\subsection{A case study with astrophysical data}
\label{sec:1es}
\begin{figure}
    \centering
    \includegraphics[width=3.5in]{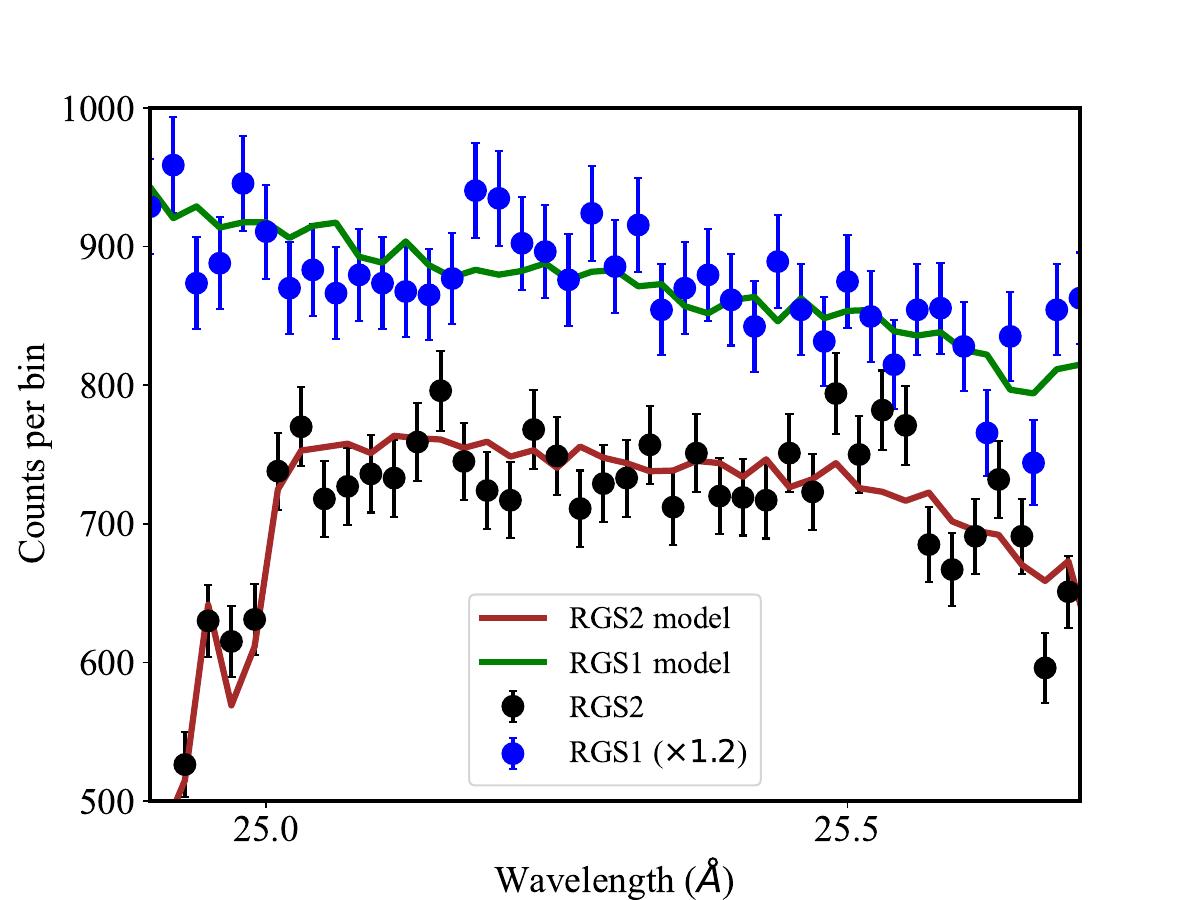}
    \caption{Zoom-in on a narrow wavelength range of the  \xmm\ spectra of the source \1es\ analyzed by \protect\cite{spence2023}. 
    The RGS1 data were shifted upward by 20\% for clarity. 
    Vertical error bars are the square root of the number of counts, 
    for illustration purposes only.}
    \label{fig:1es}
\end{figure}

The methods discussed in this section are further illustrated with the same data presented in \cite{spence2023} and \cite{bonamente2023} for the spectra of the quasar \1es. 
 The spectrum used in this case study correspond to
the data illustrated in Figure~4 and Table~2 of \cite{bonamente2023}, of which a narrow range is illustrated in Fig.~\ref{fig:1es}. 

The data are representative of a \ml\ regression with
Poisson data, in the extensive and large--count regime. 
Several data points 
have a significantly lower counts detected (i.e., below 25~\AA\ for RGS2),
 due to a decrease
in the instrument's sensitivity, or gain. Such detector inefficiencies are calibrated and
corrected \emph{in the mean} \citep[e.g.][]{tsujimoto2011}, but residual uncertainties are presumed to contribute to the larger--than--expected
variance, following the model of systematics presented in Sec.~\ref{sec:sysErrDefinition}. 
Fig.~\ref{fig:1es} illustrates how the complex parametric model generally follows the data well, without systematic
deviations which could be interpreted as a failure of the model's mean to describe the data.
Small--scale fluctuations of the model  (of order $
\leq~1$~\%) are due to changes in the sensitivity of the instrument as a function of wavelength, while large--scale trends
are a combination of wavelength--dependent sensitivity of the instrument and the genuine
spectral distribution of the source.

The data have the following parameters: $N=1,526$
 independent data  points, a model with $m=48$ free parameters, for a number of degrees of freedom $\nu=N-m=1478$, and
 a fit statistic \cmin=1862.7. The number of detected counts per bin varied across the wavelength range
 with ranges 114--1337 for RGS1 and 133--1226 for RGS2, with an average of 
 728.6 for RGS1 and 756.2 for RGS2 counts per bin; therefore these data are in the large--mean regime. 
 The one--sided $p$--value of the measured \cmin\ statistic according to the parent distribution $\chi^2(\nu)$
 is negligibly small, indicating that the null hypothesis of the Poisson data being drawn from the parent model
is formally unacceptable. Following the present model for systematic errors in the Poisson regression
 to address the larger--than--expected fit statistic, Eq.~\ref{eq:sigmaIntEst} and \ref{eq:sigmaIntConf2} yield an estimate of $\hat{f}=0.018\pm{0.02}$, meaning
 that the larger--than--expected \cmin\ statistic is consistent with a
 systematic error of approximately
 2\%, which is in fact consistent with known levels of
 detector systematics \citep{spence2023}.  

 An alternative analysis of these data consists of the \emph{assumption} of a given value for the level of systematic errors. 
 If one assumes the same point estimate $f=0.018$ based on \eqref{eq:sigmaIntEst}, it is immediate
 to see that the null hypothesis probability is 50\%: this is by design, according to \eqref{eq:muCEst}.
 A more interesting example is to test the hypothesis
 of a somewhat smaller value, say $f=0.01$ or 1\% intrinsic variance. Then \eqref{eq:muHatC} and \eqref{eq:sigmaHatC} yield a bias parameter $\hat{\mu}_C=113.2$ and an overdispersion
 parameter $\hat{\sigma}_C^2=478.1$, and
 the parent overdispersed $\chi^2$ distribution in \eqref{eq:ZDistAsymptotic} is approximated with a normal distribution
 \[
 \begin{aligned}
 \cminsyseq & \sim N(\nu+\hat{\mu}_C,2\nu+\hat{\sigma}^2_C)=N(1591.2,3434.1).
 \end{aligned}
 \]
The measured \cmin\ value corresponds to a null--hypothesis probability value of $p=0.000002$, 
indicating that this is indeed too small a value of the intrinsic variance, in accord with the estimate
above of $\hat{f}=0.018\pm{0.02}$. The same procedure can be used to test any other level of systematic errors.

\section{Discussion and Conclusions}
\label{sec:conclusions}
This paper has  presented a simple
statistical method to address the presence of systematic errors in the
regression of Poisson data, even when
there is no information on the physical source of such uncertainties. 
The method is based upon the reversal of the role played by systematic uncertainties, 
which are usually attributed to an overdispersion of
the data generating process. Within that context, overdispersed integer--count data my be modeled
with a variety of alternative distributions, such as the negative binomial \citep[see, e.g.,][]{hilbe2011,hilbe2014}.
That approach, however,
comes at the price of a more complex method of regression and its associated goodness--of--fit measure.
Instead, retaining the relatively simpler Poisson distribution for the data and attributing the overdispersion
to an intrinsic variance in the model, makes it possible to retain the  
unbiasedness  and relative simplicity of the quasi--maximum likelihood Poisson regression \citep[see][as discussed also in Sec.~\ref{sec:sigmaInt}]{gourieroux1984,gourieroux1984b}, and obtain a simple
generalization of the \gof\ statistic.

The main result of this paper is that the 
 goodness--of--fit statistic \cminsys\ for the regression of Poisson data in the presence of
 systematic errors is
the sum of two independent statistics.
The first term is the usual Poisson deviance or \emph{Cash}
statistic \cmin, with an asymptotic normal distribution in the extensive data regime 
\citep{li2024}.
The second is a corrective
term $Y$ that is also asymptotically normal, with 
simple relations given in \eqref{eq:muHatC} and \eqref{eq:sigmaHatC} for its mean and variance. 
As a result, hypothesis testing with this model for systematic errors
is simple to implement in a variety of data analysis situations, as illustrated with the numerical simulations of
Sec.~\ref{sec:cminSym} and in the case study of Sec.~\ref{sec:1es}.

This new method was motivated by the need for an accurate method to include systematics  in the regression of integer--count data, which are ubiquitous across the sciences. In addition to hypothesis testing, this method also
permits an estimate of 
the intrinsic variance from the data, when no information
is available on the level of systematic errors.
The case study based on the astrophysical spectral data \citep{spence2023} has illustrated these methods
with real--life measurements. 
The numerical simulations that we have performed for this paper show that
 the method is accurate in the Poisson large--mean limit, where we have used the chi--squared approximation
 for the distribution of the Poisson deviance $D_P$ statistic. 

\newpage
%%%%%%%%%%%%%%%%%%%%%%%%%%%%%%%%%%%%%%%%%%%%%%%%%%%%%%%%%%%%%
%%                  The Bibliography                       %%
%%                                                         %%
%%  imsart-nameyear.bst  will be used to                   %%
%%  create a .BBL file for submission.                     %%
%%                                                         %%
%%  Note that the displayed Bibliography will not          %%
%%  necessarily be rendered by Latex exactly as specified  %%
%%  in the online Instructions for Authors.                %%
%%                                                         %%
%%  MR numbers will be added by VTeX.                      %%
%%                                                         %%
%%  Use \cite{...} to cite references in text.             %%
%%                                                         %%
%%%%%%%%%%%%%%%%%%%%%%%%%%%%%%%%%%%%%%%%%%%%%%%%%%%%%%%%%%%%%
\appendix

\section{Reference distributions, compounding  {and other} formulas}
\label{app:distributions}

\subsection{The Poisson and gamma random variables}
The probability mass function of a Poisson random variable $X \sim \Poiss(\mu)$ is
\begin{equation}
    P(X=n)=\dfrac{e^{-\mu} \mu^n}{n!},\quad n=0, 1, \dots
    \label{eq:poisson}
\end{equation}
where $\mu>0$ is usually referred to as the \emph{rate} of the Poisson distribution. The mean
and variance of a $X$ are
\begin{equation}
    \E(X)=\Var(X)=\mu.
\end{equation}

The probability distribution function (pdf) of a gamma random variable $ X \sim {\rm gamma}(\alpha, r)$
is 
\begin{equation}
    f_{\gamma}(x) = \dfrac{\alpha (\alpha x)^{r-1}}{\Gamma(r)} e^{-\alpha x}, \quad x \in \mathbb{R}^+
\end{equation}
with $\Gamma$ the usual Gamma function and
\begin{comment}
\begin{equation}
    \Gamma(r)=\int_{0}^{\infty} x^{r-1}e^{-x} dx,
\end{equation}
\end{comment}
with $\alpha, r$ positive real numbers. The parameter $\alpha$ is
the \emph{rate} parameter (and $1/\alpha$ as the scale parameter), and $r$ is the \emph{shape} parameter.
The mean and variance of a $ {\rm gamma}(\alpha, r)$ random variable $X$ are
\begin{equation}
    \begin{cases}
        \E(X)=\dfrac{r}{\alpha}\\[5pt]
        \Var(X)=\dfrac{r}{\alpha^2},
    \end{cases}
\end{equation}
 with a skewness and kurtosis of respectively $\skew(X)=2/\sqrt{r}$ and  $\kurt(X)=3+6/r$.
 Other properties of the Poisson and gamma distributions can be found in most textbook on probability \citep[e.g.][]{hogg2023, siegrist}.
 
\subsection{Compounding or mixing of distributions}

Many experiments are conveniently described by 
random variables whose parameters are themselves variables following another
distribution. This \textit{compounding} of random variables is well documented
in the literature, such as the Poisson distribution with a gamma--distributed
rate parameter, i.e.,
\begin{equation}
    X \sim \Poiss(\mu)\, \quad \text{with } \mu\sim  {\rm gamma}(\alpha, r)
\end{equation}
which results in a negative binomial distribution \citep[e.g.][]{greenwood1920,bliss1953}.

Compound distributions are sometimes referred to as \textit{contagious} distribution \citep[e.g.][]{gurland1958}, based on the fact that certain models of diseases feature a (positive) correlation between the number and the probability of occurrence of certain outcomes \citep[see, e.g.,][]{greenwood1920}. More recently, this type of distributions are referred to as
\textit{mixtures}, and they are commonly used in a variety of machine learning applications,
such as the EM algorithm \citep{hastie2009, baum1970}.

General formulas for the mean and variance of a compound distribution can be given, under certain 
circumstances. Let $X$ have a distribution $F$ (i.e., Poisson), and a parameter $\theta$ of its distribution (for 
the Poisson, there is only the $\mu$ parameter) 
be distributed according to $G$; then the resulting compounded distribution $H$ is such that
\begin{equation}
\begin{aligned}
    \E_H(X)&=\E_G(\E_F(X|\theta))\\[5pt]
    \Var_H(X)&=\E_G(\Var_F(X|\theta)) + \Var_G(\E_F(X|\theta))
    \end{aligned}
    \label{eq:CompoundMoments}
\end{equation}
where the subscript highlight the relevant distribution to be used \citep[e.g.][]{mood1974}. 
In the example of the Poisson distribution with a gamma--distributed rate parameter $\mu$, which is also the mean of the Poisson
distribution, this implies 
\begin{equation}
    \begin{cases}
        E_H(X)=\mu\\[5pt]
        \Var_H(X)=\mu+\dfrac{r}{\alpha^2}=\mu+\dfrac{\mu}{\alpha}.
    \end{cases}
    \label{eq:CompoundMomentsPoissonGamma}
    \end{equation}
The results show that the Poisson distribution, compounded with a gamma--distributed rate parameter,
leads to overdispersion relative to the pure Poisson case.

\subsection{Other formulas of interest for $D_P$ }
In general, there is no guarantee that the sum of the deviations $\sum_{i=1}^N (\hat{\mu}_i - y_i)$ in a Poisson regression is equal to zero.
When an intercept is used in a GLM with the canonical logarithmic link, i.e. $\eta = \ln \mu$, then it is
immediate to see that the sum of the ML Poisson deviations does in fact equal zero. In that case,
the deviance statistic simplifies to 
\begin{equation}
D_P= 2 \sum_{i=1}^N y_i \ln(y_i/\hat{\mu}_i),
\end{equation}
and it is defined as such in certain applications \citep[e.g.][]{goodman1969}. However, 
given the emphasis on non--linear
models which may or may not have an intercept, it is convenient to retain the usual expression
\eqref{eq:DP} that applies in general.

\section{Estimates of moments of the $Y$ statistics}
\label{app:YMoments}

This appendix describes the evaluation of the moments of the $Y$ statistic,
which was defined in \eqref{eq:YStat}.
%as the additional contribution
%to the \gof\ statistic in the presence of systematic errors. 

\subsection{Definition}
The $Y$ statistic is defined by
\begin{equation}
    Y \coloneqq \cmineq(M_i)-\cmineq(\hat{\mu_i}) = 
        2 \sum_{i=1}^N (M_i-\hat{\mu}_i)-y_i \ln \left(\dfrac{M_i}{\hat{\mu}_i}\right), 
        \label{eq:YApp}
\end{equation}
where $M_i$ is a random variable with mean $\hat{\mu}_i$ and variance $\sigmaIntiEq^2$, see  \eqref{eq:Mi}.
In \eqref{eq:YApp}, $\hat{\mu}_i$ is the ML estimate of the mean of the $N$ independent 
Poisson data points $y_i$.
The goal is to 
estimate the mean $\E(Y)$ and the
variance $\Var(Y)$ as a function of the fixed parameters $\hat{\mu_i}$ and \sigmaInt, and as
a function of the data $y_i$ or its parent means $\mu_i$.

The $Y$ random variable is approximated via the usual McLaurin series for the logarithmic term,
        \begin{equation} 
        Y =  2 \sum_{i=1}^N (M_i-\hat{\mu}_i)
        -y_i\left(\dfrac{M_i-\hat{\mu}_i}{\hat{\mu}_i} \right) 
        + \dfrac{1}{2}\, y_i\left(\dfrac{M_i-\hat{\mu}_i}{\hat{\mu}_i}\right)^2 + o\left(\left(\dfrac{M_i-\hat{\mu}_i}{\hat{\mu}_i}\right)^2\right).
        \label{eq:YAppAppr}
\end{equation}
With $M_i-\hat{\mu}_i$ a zero--mean variable of variance $\sigmaIntiEq^2$, and 
$f_i=\sigmaIntRelEq \ll 1$ also by design, the expectation of the higher--order terms is much small
than the second--order term. Therefore ignoring higher--order terms provides an 
accurate approximation, and hereafter the $Y$ statistic is approximated by its first three terms
as written in \eqref{eq:YAppAppr}.

\subsection{Approximations}

It is well known that, under mild conditions on the properties of the parametric model,
the ML parameters $\hat{\theta}$ are asymptotically normally distributed,
\begin{equation}
    \sqrt{N} (\hat{\theta}-\theta) \overset{d}{\to} N(0,A^{-1})
    \label{eq:thetaHatAsymptotic}
\end{equation}
where $A$ is the Fisher information matrix, and its inverse the usual covariance or error matrix. A common 
method to estimate the parameter variances is via the Hessian estimator $\Var(\hat{\theta})$, which evaluates
the information matrix at $\theta=\hat{\theta}$ without taking the expectations \citep[e.g.][]{}. An equivalent 
$\sqrt{N}$--dependence applies to any function of the parameters, in particular to the means $\mu_i(\theta)$, whereas
the approximate \emph{delta method}, usually referred to as \emph{error propagation method} in the physical sciences, leads to
\begin{equation}
    \Var(\hat{\mu}_i) = \left. \dfrac{\partial \mu_i(\theta)^T}{\partial \theta} \right|_{\hat{\theta}} \Var(\hat{\theta}) \left. \dfrac{\partial \mu_i(\theta)}{\partial \theta^T} \right|_{\hat{\theta}}  
    \label{eq:Varmui}
\end{equation}
where $\theta$ is a column vector, and $T$ denotes its transpose \citep[for textbook references, see e.g.][]{wasserman2010, bevington2003}. For example, for the constant model considered in the Monte Carlo 
simulations of Sec.~\ref{sec:cminSym}, the partial derivatives in \eqref{eq:Varmui} are trivially equal to the unity scalar, while for the usual linear model $y=a+b\,x$ with $\theta=(a,b)$, it is $\partial \mu_i/\partial \theta = (1, x_i)$, with $x_i$ the
fixed coordinate for the $y_i$ measurement. 

For extensive data, the standard deviation of the ML in
means \eqref{eq:Varmui} therefore decreases as $\sqrt{N}$, and accordingly we approximate $\hat{\mu}_i$
as \emph{fixed} values for the purpose of estimating the moments of $Y$. It is worth pointing out that
the variability of $\hat{\mu}_i$ is duly accounted for in the distribution of the $X$ statistic in \eqref{eq:CminMi}, which
represents the usual Poisson deviance \DP\ without systematic errors. This assumption was tested with the 
Monte Carlo simulations presented in Sec.~\ref{sec:SimY}.

\subsection{Evaluation of the moments}
The assumption of independence between $M_i$ and $y_i$, by construction, 
and the assumption that $\hat{\mu}_i$ have negligible variance in the extensive data regime 
(see above) simplify considerably the
evaluation of the moments of $Y$. 
%Since the random variable $M_i$ models the presence of intrinsic systematic errors, 
%the subscript \textit{i} is used for its moments. 

The approximate estimates for the mean of $Y$ is therefore given by
\begin{equation}
    \E(Y) = 2\, \E\left[\sum_{i=1}^N \underbrace{(M_i-\hat{\mu}_i) \left(1-\dfrac{y_i}{\hat{\mu}_i}\right)}_{(A)}
        %-y_i\left(\dfrac{\mu_i-\hat{\mu}_i}{\hat{\mu}_i} \right) 
        +\underbrace{\dfrac{1}{2}\, y_i\left(\dfrac{M_i-\hat{\mu}_i}{\hat{\mu}_i}\right)^2}_{(B)}\right]
        \label{eq:EiYApp}
\end{equation}
which simply requires the given first and second--order moments for the $M_i$ variable, respectively for terms $(A)$ and $(B)$. It therefore immediately 
follows that
\begin{equation}
    \E(Y) =  \sum_{i=1}^N  \mu_i \left(\dfrac{\sigmaIntiEq }{\hat{\mu}_i} \right)^2 \simeq \sum_{i=1}^N  y_i \left(\dfrac{\sigmaIntiEq }{\hat{\mu}_i} \right)^2,
    \label{eq:muHatCApp}
\end{equation}
which is reported as \eqref{eq:muHatC}. Notice that the expectation of (A) is null regardless
of the approximation of a constant $\hat{\mu}_i$, since $\E(M_i-\hat{\mu_i})=0$. 
The second approximate equality is exact if the model $y=f(x)$ has an intercept, as do both the
constant and the linear models (and if $f_i=\sigmaIntiEq/\hat{\mu_i}$ is constant for all data points), since the ML Poisson regression in that case
satisfies $\sum (\mu_i-y_i)=0$. This approximation applies to any distribution
for $M_i$, provided the mean and variance are set according to \eqref{eq:Mi}.

For the
variance, we 
start by evaluating the moment of second order
\begin{equation}
\begin{aligned}
\E\left(Y^2\right)=& 
4\,\E \left[\left( \sum_{i=1}^N (M_i-\hat{\mu}_i) \left(1-\dfrac{y_i}{\hat{\mu}_i}\right)
        + \dfrac{1}{2}\, y_i\left(\dfrac{M_i-\hat{\mu}_i}{\hat{\mu}_i}\right)^2\right)^2 \right]\\
        =&4 \sum_{i=1}^N \E_{\Int}\left[(M_i-\hat{\mu}_i)^2\right] \cdot  \E_P\left[\left(1-\dfrac{y_i}{\hat{\mu}_i}\right)^2\right]
        + [\text{cross--product terms of (A)}]\\
        &+ \dfrac{4}{4}\sum_{i=1}^N \E_P\left(y_i^2\right) \cdot \E_{\Int}\left[\left(\dfrac{\mu_i-\hat{\mu}_i}{\hat{\mu}_i}\right)^4\right] \\
        & + \dfrac{4}{4}\underbrace{\sum_{i=1}^N \sum\limits_{\substack{j=1 \\ j\neq i}}^N 
        \E_P\left(y_i y_j\right) \cdot 
        \E_{\Int}\left[ \left(\dfrac{\mu_i-\hat{\mu}_i}{\hat{\mu}_i}\right)^2 
        \left(\dfrac{\mu_j-\hat{\mu}_j}{\hat{\mu}_j}\right)^2\right]}_{(C)}  \\
        & +[\text{cross--product terms between (A) and (B)}],
    \end{aligned}
        \label{eq:EY2}
    \end{equation}
where the subscript "$P$" indicates the Poisson distribution for the data $y_i$, and the subscript "$\Int$"  the distribution for the $M_i$ variable,
which is assumed independent from the data.
The cross--product terms of (A) and those between (A) and (B) vanish because of
independence among the $N$ terms and
the property $\E_{\Int}[\mu_i-\hat{\mu}_i]=0$. 

The cross--product terms of (B) for $i\neq j$, indicated by $C$, survive as the 
indicated double sum, which by independence evaluates to 
\[
C =  \sum_{i=1}^N \mu_i \left(\dfrac{\sigmaIntiEq}{\hat{\mu}_i}\right)^2 \cdot \sum\limits_{\substack{j=1 \\ j\neq i}}^N \mu_i \left(\dfrac{\sigmaIntjEq}{\hat{\mu}_j}\right)^2 = 
\E\left[ Y\right]^2 - \sum_{i=1}^N \mu_i^2\, f_i^4
%\left( \sum_{i=1}^N y_i \left(\dfrac{\sigmaIntiEq}{\hat{\mu}_i}\right)^2 \right)^2=\E_i(Y)^2
\]
using  \eqref{eq:muHatCApp} above, with the negative correction term arising from the $j=i$ terms
that were missing in the second sum, and indicating as usual $f_i=\sigmaIntiEq/\hat{\mu}_i$.

It therefore follows that the general expression for the approximation to the variance of
$Y$ according to the $M_i$ distribution is
\begin{equation}
\Var(Y)=
4 \sum_{i=1}^N \mu_i  f_i^2  +\sum_{i=1}^N (\mu_i^2+\mu_i) \, f_i^4 \cdot \kurt(M_i) 
%4 \sum_{i=1}^N y_i  f_i^2  +\sum_{i=1}^N y_i^2  f_i^4 \kurt(M_i) 
- \sum_{i=1}^N \mu_i^2 f_i^4, 
\label{eq:VarYApp}
\end{equation}
which depends on the kurtosis of the distribution for $M_i$.
For normal $M_i \sim N(\hat{\mu}_i,\sigmaIntiEq^2)$, $\kurt(M_i)=3$.
For a gamma distribution of the type considered in Sec.~\ref{sec:Mi},
\[
\kurt(M_i)=3 +\dfrac{6}{r}=3+6\left(\dfrac{\sigmaIntiEq}{\hat{\mu_i}}\right)^2 
= 3 +6\,f_i^2,
\]
where the excess kurtosis relative to the normal distribution provides a small correction,
since $f_i\ll1$.

A simplified expression for $\Var(Y)$ could have been obtained from \eqref{eq:EY2} with the assumption
that $y_i$ are constant, which would operationalize the variance of $Y$ as
\begin{equation}
    \Var(Y)= 4 \sum_{i=1}^N y_i\,  f_i^2  +\sum_{i=1}^N y_i^2\,  f_i^4 \cdot \kurt(M_i) 
- \sum_{i=1}^N y_i^2\, f_i^4.
\label{eq:VarYApp2}
\end{equation}
It is therefore reasonable to use either \eqref{eq:VarYApp} with the replacement of $\mu_i$ with $y_i$,
or \eqref{eq:VarYApp2}, as numerical estimators of the variance of $Y$. In the large--mean regime that is
assumed throughout,
differences between the estimators are expected to be negligible.

%% if your bibliography is in bibtex format, uncomment commands:
%\bibliographystyle{imsart-nameyear} % Style BST file
%\bibliographystyle{tfs}
\bibliographystyle{aasjournal}

%\bibliography{/home/max/proposals/max}
%\bibliography{math,max}

%% or include bibliography directly:
% \begin{thebibliography}{}
% \bibitem[\protect\citeauthoryear{???}{???}]{b1}
% \end{thebibliography}

\end{document}